\documentclass[iop]{emulateapj}

\usepackage{ulem}
\usepackage{bm}
\usepackage{color}
\usepackage{comment}
\usepackage{lineno}

\newcommand\kms{\ifmmode{\rm km\thinspace s^{-1}}\else km\thinspace s$^{-1}$\fi}

\newcommand\nar{New Astronomy Reviews}

\shortauthors{Torres}
\shorttitle{Six binaries in the Hyades}

\begin{document}
\submitted{Accepted for publication in The Astrophysical Journal}

\title{Orbits and Dynamical Masses for Six Binary Systems in the Hyades Cluster}

\author{
Guillermo Torres\altaffilmark{1},  
Gail H.\ Schaefer\altaffilmark{2}, 
Robert P.\ Stefanik\altaffilmark{1}, 
David W.\ Latham\altaffilmark{1}, 
Andrew F.\ Boden\altaffilmark{3}, 
Narsireddy Anugu\altaffilmark{2}, 
Jeremy W. Jones\altaffilmark{2}, 
Robert Klement\altaffilmark{4}, 
Stefan Kraus\altaffilmark{5}, 
Cyprien Lanthermann\altaffilmark{2}, and 
John D. Monnier\altaffilmark{6}, 
}

\altaffiltext{1}{Center for Astrophysics $\vert$ Harvard \&
  Smithsonian, 60 Garden St., Cambridge, MA 02138, USA;
  gtorres@cfa.harvard.edu}

\altaffiltext{2}{The CHARA Array of Georgia State University, Mount Wilson Observatory, Mount Wilson, CA 91203, USA}

\altaffiltext{3}{California Institute of Technology, Mail Code 11-17, 1200 East California Boulevard, Pasadena, CA 91125, USA}

\altaffiltext{4}{European Organisation for Astronomical Research in the Southern Hemisphere (ESO), Casilla 19001, Santiago 19, Chile}

\altaffiltext{5}{Astrophysics Group, Department of Physics \& Astronomy, University of Exeter, Stocker Road, Exeter, EX4 4QL, UK}

\altaffiltext{6}{Astronomy Department, University of Michigan, Ann Arbor, MI 48109, USA}

\begin{abstract}
We report long baseline interferometric observations with the
CHARA Array that resolve 6 previously known double-lined
spectroscopic binary systems in the Hyades cluster, with
orbital periods ranging from 3 to 358 days:
\object{HD~27483},
\object{HD~283882},
\object{HD~26874},
\object{HD~27149},
\object{HD~30676},
and \object{HD~28545}.
We combine those observations with new and existing radial-velocity
measurements, to infer the dynamical masses for the components
as well as the orbital parallaxes. For most stars the masses
are determined to better than 1\%. Our work significantly increases the
number of systems with mass determinations in the cluster. We find
that while current models of stellar evolution for the age
and metallicity of the Hyades are able to reproduce the overall shape of
the empirical mass-luminosity relation, they overestimate
the $V$-band fluxes by about 0.1~mag between 0.5 and 1.4~$M_{\sun}$.
The disagreement is smaller in $H$, and near zero in $K$, and
depends somewhat on the model. We also
make use of the TESS light curves to estimate rotation periods
for our targets, and detect numerous flares in one of them
(HD~283882), estimating an average flaring rate of 0.44 events per day.
\end{abstract}

\section{Introduction}
\label{sec:introduction
}

Optical interferometry has a long history of making fundamental
contributions to many areas of astrophysics \citep[see,
e.g.,][]{Quirrenbach:2001a, Quirrenbach:2001b, Monnier:2003,
Monnier:2007, Eisenhauer:2023}. Classical applications in stellar astronomy include, among many
others, the determination of orbits of close binary stars, the
measurement of angular diameters and limb darkening properties,
and the imaging of stellar surfaces.
It was just over a century ago that
the first orbit of a binary system (Capella, $\alpha$~Aur, $P =
104$~d) was measured with a 6\,m baseline Michelson
interferometer on the 100~inch telescope on Mount
Wilson \citep{Anderson:1920, Merrill:1922}. That orbit is remarkably
accurate, even by today's standards. With baselines of ever increasing
length, closer and closer binaries can be resolved, now reaching down
to periods of order a day for nearby systems. One recent example is
HD~284163 \citep{Torres:2024a}, with a period of 2.39~d.

A common scientific objective of long-baseline optical and near-infrared
interferometry is the determination of dynamical masses for
double-lined spectroscopic binaries. The mass of a star is a key ingredient for
constraining stellar evolution models, and is most useful when its precision is
better than about 3\% \citep[see, e.g.,][]{Torres:2010}.
The constraint on models becomes even stronger
when the age and metallicity of the system are also known, as is the
case for binaries that are members of well-studied clusters.  This
reduces the number of free parameters in the comparison with theory.
The example of HD~284163 mentioned above is a good illustration of
this situation. That system is part of a select group of 7
double-lined spectroscopic binaries in the Hyades cluster, whose radial
velocities (RVs) have been monitored for many years at the Center for
Astrophysics (CfA). We have also been targeting this sample recently with the
Center for High Angular Resolution Astronomy (CHARA) Array, with the
aim of resolving these systems and determining their component masses. In one
case we also make use of archival observations from the Palomar Testbed
Interferometer \citep[PTI;][]{Wallace:1998, Colavita:1999}.

The present paper reports new, high-precision dynamical mass determinations for the
remaining 6 Hyades binaries in this sample, significantly increasing
the list of systems in the Hyades with such measurements. A benefit of
astrometric-spectroscopic binaries such as these is that they also
yield a model-independent measure of the distance (orbital parallax).
We take advantage of this to improve the empirical mass-luminosity
relation in the Hyades, at both visual and near-infrared wavelengths.

The plan for the paper is as follows.
The selection and properties of our sample are explained in Section~\ref{sec:sample}.
The interferometric and spectroscopic observations of the 6 systems
are described in Sections~\ref{sec:interferometry} and
\ref{sec:spectroscopy},
respectively, and are followed in Section~\ref{sec:analysis} by a general
description of our analysis
methods that combine both types of observations to infer the masses
and orbital parallaxes. Section~\ref{sec:results}
then presents our results, with separate subsections for each system.
In Section~\ref{sec:activity} we discuss stellar activity and present
measurements of the rotation periods by making use of the light curves
from the Transiting Exoplanet Survey Satellite
mission \citep[TESS;][]{Ricker:2015}.
The mass determinations and absolute magnitudes are compared in
Section~\ref{sec:discussion} with
other observations in the Hyades, and with current stellar evolution
models. We summarize our conclusions
in Section~\ref{sec:conclusions}.

\section{Sample}
\label{sec:sample}

The list of targets for this project was drawn from a spectroscopic
survey of stars in the Hyades region, carried out for more than 40
years at the CfA. That survey has now been
essentially completed. We selected double-lined spectroscopic binary
systems with orbital periods less than a year, and with the best
determined spectroscopic orbits such that the absolute masses could
reasonably be expected to be established to better than a few percent,
given suitably precise inclination angles from the CHARA Array. The
systems were chosen to be bright enough to be accessible with CHARA
($H < 7.5$).

Of the seven FGK targets we initially selected, results for HD~284163 have already
been reported separately \citep{Torres:2024a}, as mentioned earlier. The other 6 are listed
in Table~\ref{tab:sample}, with their orbital periods as well as
coordinates, source identifiers, and parallaxes from the Gaia DR3
catalog \citep{Gaia:2022}. Other common names sometimes used in the
literature are given there as well. We
include the Gaia renormalized unit weight error (RUWE), which is an
indicator of the quality of the astrometric solution. It is typically
close to unity for sources in which a single-star model provides a
good description of the astrometric observations
\citep{Lindegren:2018}. RUWE values greater than about 1.4 can be a
sign of unmodeled binary motion \citep{Lindegren:2021a}, or other
problems with the fit.
Interestingly, Table~\ref{tab:sample} shows a trend of increasing RUWE
values with orbital period, consistent with an increase in the
amplitude of the astrometric signal one would generally
expect.\footnote{For unresolved binary systems, Gaia only measures the
  motion of the photocenter.  The amplitude of this motion will depend
  not only on the period, but also on the masses and relative
  brightness of the components.}

\setlength{\tabcolsep}{3pt}
\begin{deluxetable*}{lcccccccl}
\tablewidth{0pc}
\tablecaption{Sample of Targets\label{tab:sample}}
\tablehead{
\colhead{Target} &
\colhead{Gaia R.A.} &
\colhead{Gaia Dec.} &
\colhead{Gaia ID} &
\colhead{Period} &
\colhead{$G$} &
\colhead{$\pi_{\rm Gaia}$} &
\colhead{RUWE} &
\colhead{Aliases}
\\
\colhead{} &
\colhead{(hh:mm:ss)} &
\colhead{(dd:mm:ss)} &
\colhead{} &
\colhead{(day)} &
\colhead{(mag)} &
\colhead{(mas)} &
\colhead{} &
\colhead{}
}
\startdata
HD  27483  &  04:20:52.838  &  +13:51:51.74  &  3310615565476268032  &   3.06  &  6.05  &  $21.131 \pm 0.032$  &   1.120  &  vB 34, Han 230, HR 1358      \\ 
HD 283882  &  04:49:13.085  &  +24:48:09.38  &  147182172683187712   &   11.9  &  9.23  &  $20.327 \pm 0.033$  &   1.500  &  vB 117, BD+24~692, V808 Tau  \\ 
HD  26874  &  04:15:42.562  &  +20:49:10.71  &  49365087086285184    &   55.1  &  7.65  &  $20.307 \pm 0.044$  &   1.826  &  vB 162                       \\ 
HD  27149  &  04:18:01.965  &  +18:15:24.00  &  47620265212420096    &   75.7  &  7.37  &  $21.606 \pm 0.042$  &   1.936  &  vB 23, Han 178, V1232 Tau            \\ 
HD  30676  &  04:50:24.038  &  +17:12:09.03  &  3406103958460672768  &  224.9  &  6.96  &  $23.923 \pm 0.458$  &  19.275  &  vB 119                       \\ 
HD  28545  &  04:30:34.988  &  +15:44:01.97  &  3312631623125272448  &  358.5  &  8.67  &  $17.520 \pm 0.399$  &  21.024  &  vB 182, Han 491, Pels 61              
\enddata
\tablecomments{Coordinates, $G$-band magnitudes, parallaxes, and
  renormalized unit weight errors (RUWE) are from the Gaia DR3
  catalog. The Gaia parallaxes listed are the nominal catalog values,
  with the addition of zeropoint corrections as advocated by \cite{Lindegren:2021b}.
  See Section~\ref{sec:discussion} for corrected Gaia parallaxes for
  HD~30676 and HD~28545 that account for the orbital motion detected by
  Gaia.}
\end{deluxetable*}
\setlength{\tabcolsep}{6pt}

All of our targets have been shown to be bona fide members of the
Hyades cluster \citep[see, e.g.,][]{Griffin:1981, Griffin:2012,
  Tomkin:2003}.

\section{Interferometric Observations}
\label{sec:interferometry}

\subsection{CHARA Array Observations}

The CHARA Array is a long-baseline optical/infrared interferometer located at Mount Wilson Observatory and operated by Georgia State University \citep{tenBrummelaar:2005}. The CHARA Array combines the light from six 1\,m telescopes with baselines ranging from 34 to 331\,m. We observed the sample of Hyades binaries using the MIRC-X combiner in the $H$-band \citep{Anugu:2020} on 10 nights, and MYSTIC in the $K$-band \citep{Setterholm:2023} on the last 7 of those nights. MIRC-X and MYSTIC operate simultaneously to combine the light from all six telescopes (S1, S2, E1, E2, W1, and W2), providing spectrally dispersed visibilities on up to 15 baselines and closure phases on up to 20 triangles. Both instruments were used in their low spectral resolution mode (prism $R \sim 50$). A log of the CHARA observations is given in Table~\ref{tab:obslog}. 

Each observation consisted of recording 10 minutes of fringe data followed by a shutter sequence to measure backgrounds, foregrounds, and the ratio of light between the fringe data and the photometric channels for
each telescope. We interspersed observations of single, unresolved calibrator stars between the binary observations to calibrate the interferometric transfer function. The calibrators, adopted angular diameters \citep{Bourges:2014}, and nights on which they were observed are listed in Table~\ref{tab:cal}.  The data were reduced using the standard MIRC-X/MYSTIC pipeline\footnote{\url{https://gitlab.chara.gsu.edu/lebouquj/mircx\_pipeline.git}} \citep[version 1.3.5;][]{Anugu:2020}. We used an integration time of 2.5 min while reducing the data. The calibrated OIFITS files will be available through the Jean-Marie Mariotti Center (JMMC) Optical Interferometry Database\footnote{\url{https://www.jmmc.fr/english/tools/data-bases/oidb/}} (OIDB). As part of the reduction process, the calibrators were calibrated against each other and visually inspected; no evidence of binarity was detected in the calibrators. 

We used a binary grid search procedure\footnote{\url{http://www.chara.gsu.edu/analysis-software/binary-grid-search}} \citep{Schaefer:2016} written in the Interactive Data Language (IDL) to solve for the separation ($\rho$), position angle east of north ($\theta$) on the International Celestial Reference System (ICRS), and the flux ratio ($F_2/F_1$). The apparent sizes of the stars are unresolved by our observations, even at the longest baselines. Consequently, we adopted fixed stellar angular diameters for the primary and secondary components, and list these in Table~\ref{tab:diam}. They were estimated from preliminary masses for the components, and radii as predicted by stellar evolution models described later. During the fitting process, we divided the wavelengths in the OIFITS files by systematic correction factors of 1.0054 $\pm$ 0.0006 for MIRC-X and 1.0067 $\pm$ 0.0007 for MYSTIC (J.~D. Monnier, priv.\ comm.). The CHARA measurements for the binary positions are reported in Table~\ref{tab:CHARA}. As is customary, the 1-$\sigma$ uncertainties in the positions are defined by the major and minor axes of the error ellipse for each observation ($\sigma_{\rm maj}$, $\sigma_{\rm min}$). The orientation of the error ellipse, given by the position angle $\psi$, is dependent on the $uv$ coverage during the observation.

\setlength{\tabcolsep}{7pt}
\begin{deluxetable*}{clcl}
\tablewidth{0pc}
\tablecaption{Log for Observations at the CHARA Array\label{tab:obslog}}
\tablehead{
\colhead{ID} &
\colhead{UT Date} &
\colhead{Instrument} &
\colhead{Science Targets}
}
\startdata
01 & 2020Oct22                  & MIRC-X         & HD 283882 (5T), HD 27483 (4T, 5T), HD 26874 (5T), HD 27149 (5T)          \\ 
02 & 2020Oct23                  & MIRC-X         & HD 283882 (6T), HD 27483 (5T), HD 30676 (5T), HD 28545 (5T)              \\
03 & 2020Nov12                  & MIRC-X         & HD 27483 (6T), HD 283882 (5T), HD 26874 (5T), HD 27149 (5T)          \\
04 & 2021Oct06\tablenotemark{a} & MIRC-X/MYSTIC  & HD 283882 (4T, S1S2W1W2)                                                                 \\                                            
05 & 2021Oct22                  & MIRC-X/MYSTIC  & HD 283882 (5T) \\
06 & 2021Nov19\tablenotemark{b} & MIRC-X/MYSTIC  & HD 27483 (6T), HD 284163 (6T), HD 26874 (6T)                                              \\                               
07 & 2021Dec20                  & MIRC-X/MYSTIC  & HD 26874 (6T), HD 283882 (6T), HD 27149 (6T), HD 28545 (5T), HD 30676 (5T)    \\      
08 & 2022Oct25                  & MIRC-X/MYSTIC  & HD 27149 (5T), HD 30676 (5T), HD 26874 (5T), HD 27483 (5T), HD 28545 (5T),    \\  
   &                            &                & HD 27483 (5T) \\
09 & 2022Nov15\tablenotemark{c} & MIRC-X/MYSTIC  & HD 30676 (4T, 3T), HD 27149 (4T, 3T), HD 27483 (4T), HD 28545 (4T),    \\    
   &                            &                & HD 26874 (3T), HD 283882 (3T)  \\ 
10 & 2023Feb17                  & MIRC-X/MYSTIC  & HD 30676 (5T) 
\enddata
\tablecomments{The E1-W2-W1-S2-S1-E2 configuration was used on each night with MIRC-X in H-Prism50 mode and MYSTIC in K-Prism49 mode. Targets that were observed with all 6 telescopes are marked with ``6T'' in parentheses. After losing delay on the E1 cart in the western part of the sky, we continued observing targets with 5 telescopes (``5T'').}
\tablenotetext{a}{On UT 2021Oct06, MYSTIC was not cophased properly with MIRC-X, so only the W1-W2 and S2-S1 fringes were recorded on MYSTIC for HD 283882. No binary fit was done for this target.}
\tablenotetext{b}{On UT 2021Nov19, data on HD 27483 were recorded at a gain of 1 on MYSTIC. No binary fit was done for the MYSTIC data on this target.}
\tablenotetext{c}{On UT 2022Nov15, the S1 and S2 telescopes were offline because of a problem with the metrology signal on the S1 delay line cart and a mechanical problem with the drive bearings on the S2 telescope.}
\end{deluxetable*}

\setlength{\tabcolsep}{8pt}
\begin{deluxetable}{lllll}
\tablewidth{0pc}
\tablecaption{Adopted Calibrator Angular Diameters for the Observations at the CHARA Array\label{tab:cal}}
\tablehead{
\colhead{Calibrator} &
\colhead{UDH} &
\colhead{UDK} &
\colhead{$\sigma_{\rm UD}$} &
\colhead{Nights} \\
\colhead{} &
\colhead{(mas)} &
\colhead{(mas)} &
\colhead{(mas)} &
\colhead{Observed}
}
\startdata 
HD 17660  &  0.3053  &  0.3069  &  0.0073   & 1,2,3 \\
HD 20150  &  0.3499  &  0.3506  &  0.0127   & 9 \\
HD 23288  &  0.2277  &  0.2282  &  0.0068   & 9 \\
HD 24442  &  0.3533  &  0.3547  &  0.0084   & 5,7 \\
HD 24702  &  0.2441  &  0.2450  &  0.0057   & 5,6 \\
HD 27561  &  0.3037  &  0.3047  &  0.0076   & 8 \\
HD 27627  &  0.2727  &  0.2740  &  0.0062   & 1,2,3,8,9 \\
HD 27808  &  0.2748  &  0.2758  &  0.0066   & 1,2,3,4,5,6,7,9 \\
HD 27819  &  0.4312  &  0.4321  &  0.0400   & 10 \\
HD 28406  &  0.2766  &  0.2775  &  0.0069   & 1,2,3,6,7,8,9 \\
HD 36667  &  0.2839  &  0.2849  &  0.0069   & 5,6,8,9 
\enddata
\tablecomments{Uniform disk diameters adopted for the calibrators in the $H$-band (UDH) and $K$-band (UDK) from the JMMC Catalog of stellar diameters \citep{Bourges:2014}. The nights observed correspond to the ID column in Table~\ref{tab:obslog}. During the observations of
HD~24442 on UT 2021Oct22 and HD~24702 on UT 2021Nov19, the visibilities on the S1 baselines suffered from bad calibration caused by vibrations induced by the cable puller when the S1 delay line cart moved backwards while the star was at low elevations in the east; the S1 baselines were flagged as bad in these data files. HD~27819 was observed as a brighter calibrator on UT 2023Feb17. Although there are conflicting reports in the literature about whether it is a binary \citep[see discussion by][]{Morales:2022}, HD~27819 was used as an interferometric calibrator \citep{Boyajian:2009,Baines:2018} and has a limb-darkened diameter of $0.489 \pm 0.007$ mas measured by \citet{Salsi:2021}. The closure phases show variations less than $\pm 3^\circ$, but the MIRC-X and MYSTIC data do not give consistent results when fitting for a binary companion.}
\end{deluxetable}

\subsection{PTI Observations}

Additional, near-infrared, long-baseline interferometric measurements
of one of our targets, HD~27149, were conducted with the Palomar Testbed
Interferometer (PTI), which was a 110-m baseline $H$- and $K$-band
($\lambda\sim$1.6~$\mu$m and $\sim$2.2~$\mu$m) interferometer located
at Palomar Observatory, and decommissioned in 2008. The PTI is described in
full detail elsewhere \citep{Colavita:1999}. The instrument gave a
minimum fringe spacing of about 4~mas, making the binary orbit of the
target readily resolvable.

The PTI interferometric observable used for these measurements is the
fringe contrast or ``visibility" (specifically, the power-normalized
visibility modulus squared, $V^2$) of the observed brightness
distribution on the sky. The measurements we obtained were made in
the $K$ band, and are given in
Table~\ref{tab:PTI}.  HD~27149 was typically observed in conjunction with
two calibration objects, and each observation (or scan) was
approximately 130~s long.  As in previous publications, PTI $V^2$
data reduction and calibration follows standard procedures described
by \citet{Colavita:2003} and \cite{Boden:1998}, respectively.  For this
analysis, we used \objectname[HD 27397]{HD~27397} and \objectname[HD
27459]{HD~27459} as our calibration sources, with adopted uniform-disk
angular diameters of $0.40 \pm 0.07$ and $0.38 \pm 0.06$~mas, respectively. At these
diameter estimates, the PTI baselines do not significantly resolve
these sources.

\setlength{\tabcolsep}{8pt}
\begin{deluxetable}{lll}
\tablewidth{10pc}
\tablecaption{Adopted Angular Diameters for Primary and Secondary Stars in the Hyades Binaries\label{tab:diam}}
\tablehead{
\colhead{Binary} &
\colhead{UD$_1$} &
\colhead{UD$_2$} \\
\colhead{} &
\colhead{(mas)} &
\colhead{(mas)}
}
\startdata 
HD 27483   &  0.265  &  0.257 \\
HD 283882  &  0.134  &  0.127 \\
HD 26874   &  0.186  &  0.163 \\
HD 27149   &  0.199  &  0.179 \\
HD 30676   &  0.156  &  0.126 \\
HD 28545   &  0.159  &  0.118 
\enddata
\end{deluxetable}

\setlength{\tabcolsep}{5pt}
\begin{deluxetable*}{lccccccccl}
\tablecaption{CHARA Measurements for our Targets \label{tab:CHARA}}
\tablehead{
\colhead{Target} &
\colhead{UT Date} &
\colhead{HJD} &
\colhead{$\rho$} &
\colhead{$\theta$} &
\colhead{$\sigma_{\rm maj}$} &
\colhead{$\sigma_{\rm min}$} &
\colhead{$\psi$} &
\colhead{$F_2/F_1$} &
\colhead{Instrument}
\\
\colhead{} &
\colhead{} &
\colhead{(2,400,000+)} &
\colhead{(mas)} &
\colhead{(degree)} &
\colhead{(mas)} &
\colhead{(mas)} &
\colhead{(degree)} &
\colhead{} &
\colhead{}
}
\startdata
HD 27483  &  2020Oct22  &  59144.949  &  1.1649  &   27.30  &  0.0014  &  0.0012  &  174.72  &  0.929 &  MIRC-X \\
HD 27483  &  2020Oct22  &  59144.966  &  1.1548  &   29.06  &  0.0016  &  0.0012  &  167.48  &  0.928 &  MIRC-X \\
HD 27483  &  2020Oct23  &  59145.976  &  1.1069  &  163.43  &  0.0067  &  0.0024  &   59.20  &  0.928 &  MIRC-X \\
HD 27483  &  2020Nov12  &  59165.802  &  1.0672  &  337.60  &  0.0013  &  0.0004  &  118.77  &  0.933 &  MIRC-X \\
HD 27483  &  2021Nov19  &  59537.768  &  1.1955  &  184.25  &  0.0080  &  0.0072  &  101.50  &  0.943 &  MIRC-X 
\enddata

\tablecomments{The columns $\sigma_{\rm maj}$ and $\sigma_{\rm
      min}$ represent the major and minor axes of the 1-$\sigma$ error
    ellipse for each measurement, and $\psi$ gives the orientation of
    the major axis relative to the direction to the north.  All
  position angles are referred to the International Celestial
  Reference Frame (effectively J2000). Formal uncertainties for the
  flux ratios $F_2/F_1$ in the $H$ or $K$ bands are not reported, as
  they are typically unrealistically small. A more representative
  value for the uncertainty is given by the scatter of the
  measurements from different epochs (see
  Section~\ref{sec:discussion}).  (This table is available in its
  entirety in machine-readable form.)}

\end{deluxetable*}
\setlength{\tabcolsep}{6pt}

\setlength{\tabcolsep}{5pt}
\begin{deluxetable*}{lcccccc}
\tablecaption{PTI $K$-band Squared Visibility Measurements for HD 27149 \label{tab:PTI}}
\tablehead{
\colhead{UT Date} &
\colhead{HJD} &
\colhead{$\langle\lambda\rangle$} &
\colhead{$V^2$} &
\colhead{$\sigma_{V^2}$} &
\colhead{$u$} &
\colhead{$v$}
\\
\colhead{} &
\colhead{(2,400,000+)} &
\colhead{($\mu$m)} &
\colhead{} &
\colhead{} &
\colhead{(m)} &
\colhead{(m)}
}
\startdata
2000Sep13 &   51801.0255 &  2.2306  &  0.809  &  0.074  &  $-36.4936$  &  $-98.9544$ \\
2000Sep13 &   51801.0331 &  2.2266  &  0.828  &  0.067  &  $-33.5676$  &  $-99.4827$ \\
2000Sep13 &   51801.0402 &  2.2402  &  1.028  &  0.167  &  $-30.7831$  &  $-99.9330$ \\
2000Oct01 &   51818.9080 &  2.2320  &  0.586  &  0.570  &  $-58.1778$  &  $-92.4689$ \\
2000Oct01 &   51818.9100 &  2.2290  &  0.771  &  0.305  &  $-57.6839$  &  $-92.6953$ 
\enddata
\tablecomments{$\langle\lambda\rangle$ is the flux-weighted center-band
wavelength. The last two columns represent the projected baselines.
  (This table is available in its entirety in machine-readable form.)}
\end{deluxetable*}
\setlength{\tabcolsep}{6pt}

\section{Spectroscopic Observations}
\label{sec:spectroscopy}

Our sample of objects in Table~\ref{tab:sample} has been observed
spectroscopically at the CfA for a decade or more, as part of a
larger program to monitor the radial velocities of several hundred
stars in the Hyades region.  The vast majority of
the observations for this paper were gathered with two nearly
identical echelle instruments \citep[Digital
  Speedometers;][]{Latham:1992}, which operated until 2010.  One was
attached to the 1.5m Wyeth reflector at the (now closed) Oak Ridge
Observatory, in the town of Harvard (Massachusetts, USA), and the
other to the 1.5m Tillinghast reflector at the Fred L.\ Whipple
Observatory (Mount Hopkins, Arizona, USA). These instruments were
equipped with intensified photon-counting Reticon detectors that recorded a single
echelle order 45~\AA\ wide, centered on the \ion{Mg}{1}~b triplet near
5187~\AA.  The resolving power was $R \approx 35,\!000$, corresponding
to 8.5~\kms.

A few observations for one of our targets (HD~28545) were gathered
more recently with the Tillinghast Reflector Echelle Spectrograph
\citep[TRES;][]{Szentgyorgyi:2007, Furesz:2008}, a bench-mounted,
fiber-fed instrument on the 1.5m telescope in Arizona. The TRES CCD
detector records 51 orders over the 3800--9100~\AA\ range, at a
resolving power of $R \approx 44,\!000$ (6.8~\kms).

For the Digital Speedometers, the zeropoint of the radial velocities
(RVs) was monitored by taking sky exposures in the evening and the
morning, which were used to calculate and apply small run-to-run
corrections to the raw velocities in order to place them on a common
system \citep[see][]{Latham:1992}. These corrections were typically
under 2~\kms. This native CfA system is slightly offset from the IAU
reference frame by 0.14~\kms\ \citep{Stefanik:1999}, as determined
from observations of minor planets in the solar system. In order to
remove this shift, we adjusted the velocities by adding +0.14~\kms.
For TRES, changes in the velocity zeropoint were monitored with
observations of several IAU standard stars, and asteroid observations
were then employed to transfer the raw velocities to an absolute
system, as with the Digital Speedometers.

The spectra of all our targets are double-lined. Radial velocities
were measured using TODCOR, which is a two-dimensional
cross-correlation algorithm \citep{Zucker:1994}. Templates were
selected from a large library of synthetic spectra based on model
atmospheres by R.\ L.\ Kurucz, and a line list manually tuned to
better match real stars \citep[see][]{Nordstrom:1994, Latham:2002}.
These templates cover the region centered on the
\ion{Mg}{1}\,b triplet, which captures most of the velocity information.
The surface gravity for the templates was held at
$\log g = 4.5$, which is appropriate for our objects, and we assumed
solar metallicity, which is sufficiently close to the Hyades
composition for our purposes \citep[${\rm [Fe/H]} =
  +0.18$;][]{Dutra-Ferreira:2016}. To determine the optimal effective
temperature and line broadening for each binary component ($T_{\rm
  eff}$ and $v \sin i$), we ran grids of cross-correlations over broad
ranges in those parameters, as described by \cite{Torres:2002}.  We
then selected the templates producing the highest correlation
coefficient averaged over all exposures.  The adopted template
parameters are reported in Section~\ref{sec:results}. We note that
while rotation may be the dominant line broadening mechanism for these
stars, what we refer to here as $v \sin i$, for short,
includes all other broadening mechanisms, such as macroturbulence
(beyond the value of $\zeta_{\rm RT} = 1~\kms$ already built into our
templates).

Experience has shown that the limited number of lines in the narrow
wavelength range of the Digital Speedometers can sometimes introduce
subtle biases in the RVs. These errors are caused by spectral lines of
the binary components shifting in and out of the spectral window in opposite directions, depending
on both the orbital phase and the projected velocity of the
Earth relative to the solar system barycenter. We evaluated and
corrected for these effects in each of our systems through numerical
simulations. We refer the reader to \cite{Latham:1996} and
\cite{Torres:1997a} for further details of that analysis. The measured radial velocities
and formal uncertainties for all our targets are presented together in
Table~\ref{tab:rvs}. The use of TODCOR allows us to extract for each
binary an estimate of the spectroscopic flux ratio between the
components \citep[see][]{Zucker:1994}, at the mean wavelength of our
observations ($\sim$5187~\AA). We report these values in
Section~\ref{sec:results}, and use them later to infer the individual
brightness of each component.

\setlength{\tabcolsep}{6pt}  
\begin{deluxetable*}{lccccc}
\tablewidth{0pc}
\tablecaption{CfA Radial Velocities for our Targets \label{tab:rvs}}
\tablehead{
\colhead{Target Name} &
\colhead{HJD} &
\colhead{$RV_1$} &
\colhead{$\sigma_1$} &
\colhead{$RV_2$} &
\colhead{$\sigma_2$}
\\
\colhead{} &
\colhead{(2,400,000+)} &
\colhead{(\kms)} &
\colhead{(\kms)} &
\colhead{(\kms)} &
\colhead{(\kms)}
}
\startdata
HD 27483  &  44560.8138  &  \phn86.21           &  0.62 &  $-$10.82\phs &  0.52 \\
HD 27483  &  44591.7209  &  \phn41.95           &  0.54 &   33.64     &  0.45 \\
HD 27483  &  44627.6614  &  111.22              &  1.11 &  $-$34.88\phs &  0.93 \\
HD 27483  &  44629.6680  &  \phn\phn$-$6.49\phs &  0.72 &   83.74     &  0.60 \\
HD 27483  &  44887.8474  &  107.26              &  0.74 &  $-$34.08\phs &  0.62
\enddata
\tablecomments{The velocities in this table include all adjustments
described in the text. (This table is available in its entirety in machine-readable form.)}
\end{deluxetable*}
\setlength{\tabcolsep}{6pt}  

In addition to our own velocities, the analyses described in the next
section made use of RV observations from other sources that extend our
time baseline or supplement the phase coverage, and are of sufficient
quality to make them useful. These included measurements from the
extensive Hyades program conducted by Roger Griffin
\citep{Griffin:1978, Griffin:1981, Griffin:2012}, as well as
velocities with the CORAVEL spectrometer as reported by
\cite{Mermilliod:2009}. In several cases we also used velocity
measurements taken from the public archive of observations
gathered with the Elodie spectrograph\footnote{\url{http://atlas.obs-hp.fr/elodie/}},
which are typically of high precision. Other RV sources are described below.

\section{Orbital Analysis}
\label{sec:analysis}

The CHARA observations and RV measurements for each system were used
together to solve for the astrometric and spectroscopic orbital
elements simultaneously. The usual spectroscopic elements are the
period ($P$), center-of-mass velocity ($\gamma$), primary and
secondary velocity semiamplitudes ($K_1$, $K_2$), the eccentricity
($e$) and argument of periastron for the primary ($\omega_1$), expressed for this work as
$\sqrt{e}\cos\omega_1$ and $\sqrt{e}\sin\omega_1$, and a reference
time of periastron passage ($T_{\rm peri}$). The purely astrometric
elements are the angular semimajor axis ($a^{\prime\prime}$), the cosine of the
orbital inclination angle relative to the line of sight ($\cos i$), and the
position angle of the ascending node for J2000.0 ($\Omega$). 
In order to account for the possibility of systematic shifts in
the velocities from outside sources relative to our own, we typically
also solved for separate RV offsets ($\Delta$) for each of those
sources, together with the other free parameters of our model. These offsets should
be added to the corresponding velocities in order to refer them to the
reference frame of the CfA velocities. For the PTI observations of
HD~27149, we included the $K$-band flux ratio as an additional
free parameter.

Our solutions were performed within a Markov Chain Monte Carlo
framework, using the {\sc
  emcee}\footnote{\url{https://emcee.readthedocs.io/en/stable/index.html}}
package of \cite{Foreman-Mackey:2013}. We applied uniform priors over
suitable ranges for all of the above adjustable parameters.
Convergence was verified by visual inspection of the chains. We also
required a Gelman-Rubin statistic of 1.05 or smaller \citep{Gelman:1992}.

As internal measurement errors are not always accurate, but are
important for the proper weighting of the various datasets relative to
one another, we included additional free parameters in our analysis to
represent multiplicative error scaling factors, $f$, for all
uncertainties. We did this for each RV dataset, separately for the
primary and secondary components, as well as for the CHARA
observations and the PTI squared visibilities. These factors were solved for
simultaneously and self-consistently with the other free parameters
\citep[see][]{Gregory:2005}, using log-normal priors.

\section{Results}
\label{sec:results}

The subsections that follow present a summary of the available
spectroscopic data, and the particulars of the orbital
analysis for each of our six Hyades binaries, in order of increasing
orbital period, as in Table~\ref{tab:sample}. Aside from the component masses, the
astrometric-spectroscopic solutions yield the orbital parallax, which
we use later to compute the absolute brightness of the components.

\subsection{HD 27483}
\label{sec:hd27483}

This Hyades object has been known to be a spectroscopic binary for
more than eight decades \citep{Christie:1938, Young:1939}. The first
double-lined orbital solution, with a period of 3.06~days, was
reported by \cite{Northcott:1952}, on the basis of 67 photographic RV
measurements made between 1933 and 1951 at the Dominion Astrophysical
Observatory. Additional velocities with the CORAVEL instrument have
been reported by \cite{Mayor:1987} and \cite{Mermilliod:2009}, the
latter ones largely superseding the former ones, after a zeropoint
adjustment and changes to the uncertainties. As described in more
detail by \cite{Griffin:2012}, these two recent datasets are not exactly the
same, and the earlier publication has the identities of the primary and
secondary reversed.  Confusingly, the 2009 paper has the wrong
assignments in several instances. \cite{Mayor:1987} reported RVs at 5
epochs that are not included in the \cite{Mermilliod:2009} set. Based on
a comparison of the observations in common, we have scaled the original
uncertainties of those 5 measurements by a factor 1.6, and shifted
them to the \cite{Mermilliod:2009} system by applying the same average
adjustment of +0.34~\kms\ used by those authors. Altogether, there
are 20 CORAVEL observations of the primary and 19 of the secondary
taken between 1979 and 1993, excluding two that were gathered at times
when the lines of the two stars were severely blended. We have also
used 25 pairs of velocities (1986--2010) by \cite{Griffin:2012}, with
the relative weights assigned by that author, and adopting his error of
0.70~\kms\ for an observation of unit weight.  Compared to these more
recent measurements, the ones by \cite{Northcott:1952} are so much
poorer that they provide no benefit for our analysis, not even for
improving the period. Consequently, we have not used them.

Our own spectroscopic contribution consists of 114 observations
made with the Digital Speedometers between 1980 and 2001,
with signal-to-noise ratios (SNRs)
ranging from 11 to 81 per resolution element. The synthetic templates
adopted for the cross-correlations with TODCOR have temperatures of
6500~K for both components, and rotational broadenings of 20 and
12~\kms\ for the primary and secondary. The velocities have been
listed earlier in Table~\ref{tab:rvs}. We determined a spectroscopic
flux ratio between the components of $(F_2/F_1)_{\rm sp} = 0.854 \pm 0.027$, at
the mean wavelength of our observations.

HD~27483 was also observed with the Palomar Testbed Interferometer (PTI) by
\cite{Konacki:2004}, who presented a spectroscopic-astrometric orbital
solution based on the RV measurements of \cite{Mayor:1987}. However,
as the authors reported, the binary was only partially resolved,
causing a strong degeneracy between the semimajor axis and the
brightness ratio. This compelled them to assume an arbitrary brightness ratio of
unity. Their mass determinations have formal uncertainties approaching
10\%. We experimented with incorporating those PTI visibilities into our
analysis, but found no improvement in any of the elements compared to
using only the CHARA observations. Consequently, those PTI observations
have not been used here.

\begin{figure}
\epsscale{1.15}
\plotone{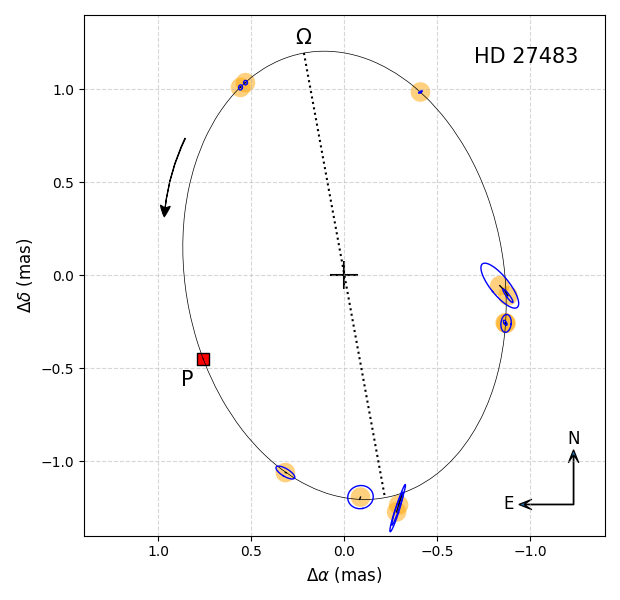}
\figcaption{Astrometric measurements for HD~27483, with our model for
  the orbit. Each observation is highlighted with an orange circle
  for clarity, and is shown with its error ellipse. 
  Short line segments connect the measured positions with the
  predicted ones from our model. The line of nodes is represented with a
  dotted line, and the ascending node is indicated with the ``$\Omega$"
  symbol. Periastron is marked with the red square labeled ``P''.
  \label{fig:CHARA_hd27483}}
\end{figure}

\begin{figure}
\epsscale{1.18} \plotone{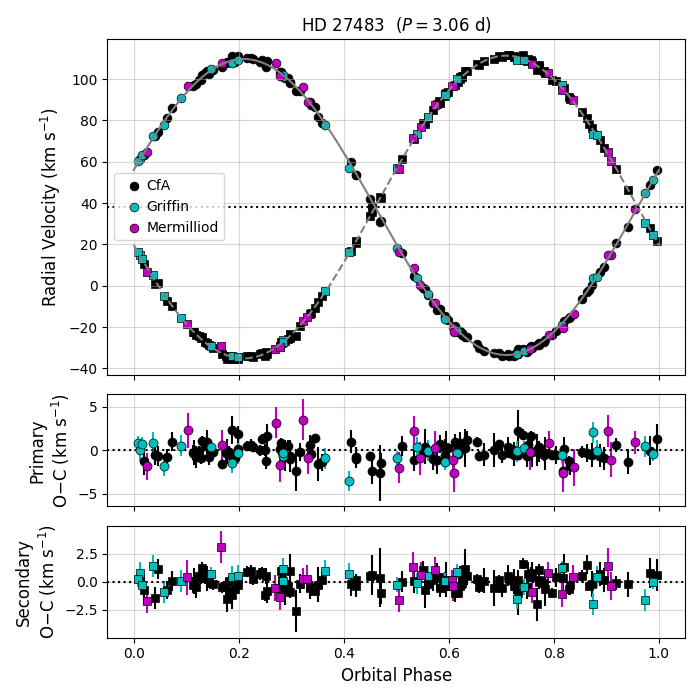}
\figcaption{RV
  measurements for HD~27483 from the CfA and other literature sources,
  along with our adopted model for the spectroscopic orbit. Phase
  zero corresponds to the formal time of periastron passage listed
  in Table~\ref{tab:hd27483_results}. The dotted
  line represents the center-of-mass velocity. Residuals with the
  corresponding error bars are shown in the lower panels. Some error
  bars are smaller than the symbol size.\label{fig:RV_hd27483}}
\end{figure}

In addition to the standard orbital elements and error inflation
factors for each data set described earlier, we allowed for systematic
offsets ($\Delta_{\rm G}$, $\Delta_{\rm M}$) for the \cite{Griffin:2012}
and \cite{Mermilliod:2009} velocities relative to ours.
Initial spectroscopic-only solutions indicated a negligible
eccentricity, and in such cases the time of periastron passage is
poorly determined. To avoid this indeterminacy in our analysis,
instead of $T_{\rm peri}$ we solved for the time of nodal passage
$T_{\rm node}$, which is always well determined. It corresponds to the
time of maximum primary velocity.  Table~\ref{tab:hd27483_results}
presents the results, along with other derived properties including
the masses. The eccentricity is not statistically significant, and we
infer a 3$\sigma$ upper limit of $e = 0.002$. The astrometric orbit
with our CHARA measurements is shown in
Figure~\ref{fig:CHARA_hd27483}, and the radial velocities can be seen
in Figure~\ref{fig:RV_hd27483}.


\setlength{\tabcolsep}{12pt}
\begin{deluxetable}{lc}
\tablewidth{0pc}
\tablecaption{Results of our Orbital Analysis for HD~27483
\label{tab:hd27483_results}}
\tablehead{
\colhead{~~~~~~~~~~~Parameter~~~~~~~~~~~} &
\colhead{Value}
}
\startdata
 $P$ (day)                          & $3.05911946^{+0.00000041}_{-0.00000041}$ \\ [1ex]
 $T_{\rm node}$ (HJD$-$2,400,000)   & $49207.19931^{+0.00053}_{-0.00053}$      \\ [1ex]
 $a^{\prime\prime}$ (mas)           & $1.2152^{+0.0021}_{-0.0021}$             \\ [1ex]
 $\sqrt{e}\cos\omega_1$             & $+0.005^{+0.017}_{-0.018}$               \\ [1ex]
 $\sqrt{e}\sin\omega_1$             & $-0.019^{+0.021}_{-0.013}$               \\ [1ex]
 $\cos i$                           & $0.7042^{+0.0017}_{-0.0017}$             \\ [1ex]
 $\Omega$ (degree)                  & $10.44^{+0.16}_{-0.16}$                  \\ [1ex]
 $\gamma$ (\kms)                    & $+38.086^{+0.049}_{-0.049}$              \\ [1ex]
 $K_1$ (\kms)                       & $71.642^{+0.086}_{-0.086}$               \\ [1ex]
 $K_2$ (\kms)                       & $73.273^{+0.070}_{-0.070}$               \\ [1ex]
 $\Delta_{\rm G}$ (\kms)            & $-0.27^{+0.14}_{-0.14}$                  \\ [1ex]
 $\Delta_{\rm M}$ (\kms)            & $-1.29^{+0.23}_{-0.23}$                  \\ [1ex]
 $f_{\rm CHARA}$                    & $2.94^{+0.62}_{-0.41}$                   \\ [1ex]
 $f_{\rm CfA,1}$ and $\sigma_1$ (\kms) & $0.886^{+0.065}_{-0.055}$, 0.80          \\ [1ex]
 $f_{\rm CfA,2}$ and $\sigma_2$ (\kms) & $0.867^{+0.062}_{-0.053}$, 0.66          \\ [1ex]
 $f_{\rm G,1}$ and $\sigma_1$ (\kms)   & $1.19^{+0.22}_{-0.14}$, 0.99             \\ [1ex]
 $f_{\rm G,2}$ and $\sigma_2$ (\kms)   & $0.97^{+0.18}_{-0.12}$, 0.80             \\ [1ex]
 $f_{\rm M,1}$ and $\sigma_1$ (\kms)   & $1.63^{+0.32}_{-0.21}$, 1.79             \\ [1ex]
 $f_{\rm M,2}$ and $\sigma_2$ (\kms)   & $1.14^{+0.24}_{-0.16}$, 1.10             \\ [1ex]
\hline \\ [-1.5ex]
\multicolumn{2}{c}{Derived quantities} \\ [1ex]
\hline \\ [-1.5ex]
 $e$                                & $0.00062^{+0.00072}_{-0.00041}$          \\ [1ex]
 $\omega_1$ (degree)                & $280^{+36}_{-128}$                       \\ [1ex]
 $i$ (degree)                       & $45.24^{+0.13}_{-0.13}$                  \\ [1ex]
 $M_1$ ($M_{\sun}$)                 & $1.363^{+0.010}_{-0.010}$             \\ [1ex]
 $M_2$ ($M_{\sun}$)                 & $1.3323^{+0.0099}_{-0.0099}$             \\ [1ex]
 $q \equiv M_2/M_1$                 & $0.9777^{+0.0015}_{-0.0015}$             \\ [1ex]
 $a$ ($R_{\sun}$)                   & $12.341^{+0.030}_{-0.030}$               \\ [1ex]
 $\pi_{\rm orb}$ (mas)              & $21.174^{+0.072}_{-0.073}$               \\ [1ex]
 Distance (pc)                      & $47.23^{+0.16}_{-0.16}$                  \\ [1ex]
 $T_{\rm peri}$ (HJD$-$2,400,000)   & $49209.58^{+0.31}_{-1.09}$      
\enddata

\tablecomments{$f_{\rm CfA,1}$ and $f_{\rm CfA,2}$ are the scale
  factors for the internal errors of the CfA RV velocities of the
  primary and secondary. A similar notation is used for the RVs of
  \cite{Griffin:2012} and \cite{Mermilliod:2009}. Values following
  these scale factors on the same line are the weighted rms residuals,
  after application of the scale factors.}

\end{deluxetable}
\setlength{\tabcolsep}{6pt}

HD~27483 has a distant white dwarf companion, first detected
spectroscopically by \cite{Bohm-Vitense:1993} from excess flux in an
IUE spectrum. The Gaia DR3 catalog has no entry for this
companion. It was spatially resolved using HST by
\cite{Barstow:2001}, who found it to be about 1\farcs3 due north, and
1.96 magnitudes fainter in the UV. Additional measurements of the
relative position and brightness with HST and ground-based adaptive
optics were reported by \cite{Zhang:2023}.  These authors combined
those constraints with the astrometric acceleration of the primary
from the proper motion difference between Gaia and Hipparcos, to infer
a mass for the white dwarf of $0.798^{+0.10}_{-0.04}~M_{\sun}$, an
orbital period of $184^{+65}_{-30}$~yr, and an orbital inclination of
$30^{+13}_{-15}$~deg. The system is therefore triple. This is not
unexpected, as \cite{Tokovinin:2006} have shown that more than 95\% of
spectroscopic binaries with periods under about 3 days have additional
components. \cite{Zhang:2023} noted that the orbit of the white dwarf
appears reasonably well aligned with that of the inner binary. They based
this conclusion on an adopted inclination angle for the inner binary
of $45\fdg1 \pm 1\fdg7$ from \cite{Konacki:2004}, which is similar to ours.

\subsection{HD 283882}
\label{sec:hd283882}

Double lines in the spectra of HD~283882 were first mentioned by \cite{Wilson:1948},
and later also by \cite{Young:1974}, who provided the first 5 RV measures
of each component made in 1971. \cite{Griffin:1978} observed it more
extensively, and derived a spectroscopic orbit with a period of
11.9~days. That set of observations was augmented (and the date of the
last one corrected) by \cite{Griffin:2012}, giving much more complete
phase coverage in the rather eccentric orbit. After rejecting several
observations at strongly blended phases, and a few others that
\cite{Griffin:2012} also considered unreliable, there are 60
measurements of the primary and 62 of the secondary made between 1973
and 2010, which we used for our analysis below. We adopted the relative
weighting recommended by Griffin, along with his stated uncertainty of
0.55~\kms\ for an observation of unit weight.

Forty observations for HD~283882 were collected at the CfA with the
Digital Speedometers between 1985 and 1995, of which one giving
abnormally large velocity residuals for both components was rejected.
Signal-to-noise ratios are relatively low in this case, and range
between 8 and 17 per resolution element.  The cross-correlation
templates that were found to give the best results have temperatures
of 5000 and 4750~K for the primary and secondary, and rotational
broadenings of 6~\kms\ for both stars. We estimate the spectroscopic flux
ratio between the components to be $(F_2/F_1)_{\rm sp} = 0.524 \pm 0.045$.

There are two other smaller sets of velocities available
that were also used in our modeling of HD~283882. One consists of 8
primary and 6 secondary velocities by \cite{Mermilliod:2009}, made with
the CORAVEL instrument (1978--1981), with their corresponding formal
uncertainties.  The other is a set of 6 archival velocity measurements
for both stars from 2003--2005, which we retrieved from the public
Elodie archive. For these, we adopted arbitrary initial uncertainties
of 0.5~\kms, later adjusted by iterations in the analysis below. Even
though the 5 pairs of old measurements by \cite{Young:1974} are quite
consistent with the orbit model presented in this section, we chose
not to incorporate them because of their larger scatter.

\begin{figure}
\epsscale{1.15}
\plotone{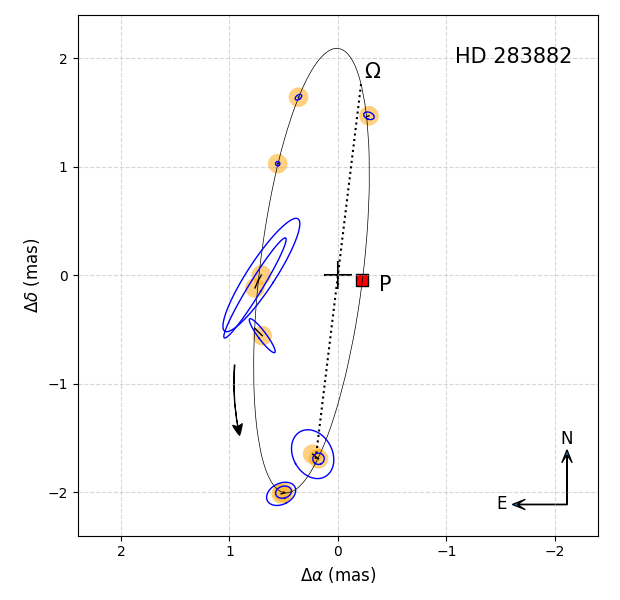}
\figcaption{Astrometric measurements for HD~283882, as in
  Figure~\ref{fig:CHARA_hd27483}.
  \label{fig:CHARA_hd283882}}
\end{figure}

\begin{figure}
\epsscale{1.18} \plotone{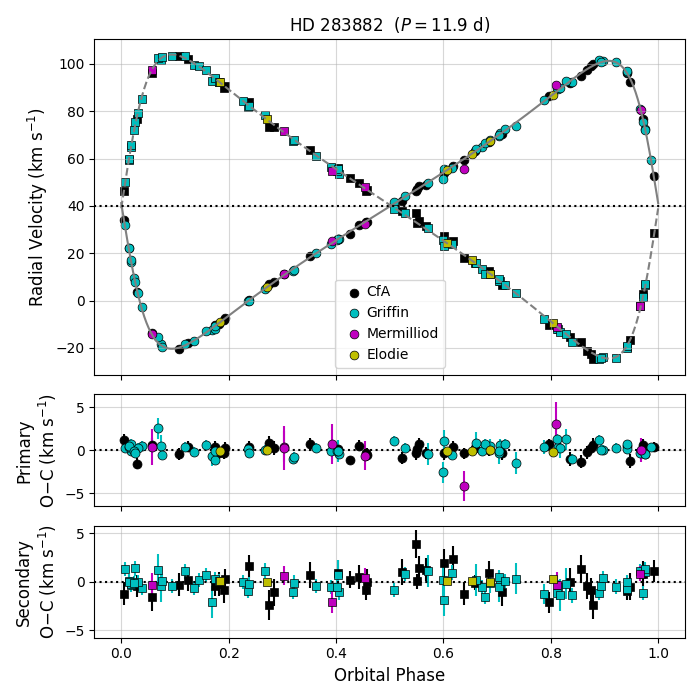} \figcaption{Same as
  Figure~\ref{fig:RV_hd27483}, for HD~283882.\label{fig:RV_hd283882}}
\end{figure}

The joint solution with the velocities and our CHARA observations is
presented in Table~\ref{tab:hd283882_results}, and includes additional
free parameters to account for possible systematic offsets between the
Griffin, Mermilliod, and Elodie velocities and our own ($\Delta_{\rm
  G}$, $\Delta_{\rm M}$, $\Delta_{\rm E}$).  The model and the
observations are shown in Figures~\ref{fig:CHARA_hd283882} and
\ref{fig:RV_hd283882}.

\setlength{\tabcolsep}{12pt}
\begin{deluxetable}{lc}
\tablewidth{0pc}
\tablecaption{Results of our Orbital Analysis for HD~283882 \label{tab:hd283882_results}}
\tablehead{
\colhead{~~~~~~~~~~~Parameter~~~~~~~~~~~} &
\colhead{Value}
}
\startdata
 $P$ (day)                          & $11.928690^{+0.0000058}_{-0.0000057}$  \\ [1ex]
 $T_{\rm peri}$ (HJD$-$2,400,000)   & $47820.9470^{+0.0022}_{-0.0022}$       \\ [1ex]
 $a^{\prime\prime}$ (mas)           & $2.4144^{+0.0076}_{-0.0077}$           \\ [1ex]
 $\sqrt{e}\cos\omega_1$             & $+0.0098^{+0.0013}_{-0.0015}$          \\ [1ex]
 $\sqrt{e}\sin\omega_1$             & $+0.71937^{+0.00054}_{-0.00057}$       \\ [1ex]
 $\cos i$                           & $0.1947^{+0.0018}_{-0.0018}$           \\ [1ex]
 $\Omega$ (degree)                  & $353.09^{+0.21}_{-0.21}$               \\ [1ex]
 $\gamma$ (\kms)                    & $+40.099^{+0.092}_{-0.091}$            \\ [1ex]
 $K_1$ (\kms)                       & $60.856^{+0.086}_{-0.078}$             \\ [1ex]
 $K_2$ (\kms)                       & $64.25^{+0.11}_{-0.10}$              \\ [1ex]
 $\Delta_{\rm G}$ (\kms)            & $-0.75^{+0.11}_{-0.11}$                \\ [1ex]
 $\Delta_{\rm M}$ (\kms)            & $+0.22^{+0.39}_{-0.36}$                \\ [1ex]
 $\Delta_{\rm E}$ (\kms)            & $+0.01^{+0.11}_{-0.11}$                \\ [1ex]
 $f_{\rm CHARA}$                    & $3.26^{+0.78}_{-0.51}$                 \\ [1ex]
 $f_{\rm CfA,1}$ and $\sigma_1$ (\kms) & $1.08^{+0.15}_{-0.11}$, 0.64           \\ [1ex]
 $f_{\rm CfA,2}$ and $\sigma_2$ (\kms) & $0.961^{+0.127}_{-0.092}$, 1.16        \\ [1ex]
 $f_{\rm G,1}$ and $\sigma_1$ (\kms)   & $1.006^{+0.111}_{-0.086}$, 0.60        \\ [1ex]
 $f_{\rm G,2}$ and $\sigma_2$ (\kms)   & $1.070^{+0.114}_{-0.089}$, 0.79        \\ [1ex]
 $f_{\rm M,1}$ and $\sigma_1$ (\kms)   & $2.92^{+1.06}_{-0.57}$, 1.80           \\ [1ex]
 $f_{\rm M,2}$ and $\sigma_2$ (\kms)   & $1.33^{+0.65}_{-0.30}$, 0.92           \\ [1ex]
 $f_{\rm E,1}$ and $\sigma_1$ (\kms)   & $0.234^{+0.166}_{-0.074}$, 0.104       \\ [1ex]
 $f_{\rm E,2}$ and $\sigma_2$ (\kms)   & $0.283^{+0.209}_{-0.092}$, 0.124       \\ [1ex]
\hline \\ [-1.5ex]
\multicolumn{2}{c}{Derived quantities} \\ [1ex]
\hline \\ [-1.5ex]
 $e$                                & $0.51759^{+0.00076}_{-0.00081}$        \\ [1ex]
 $\omega_1$ (degree)                & $89.22^{+0.12}_{-0.11}$                \\ [1ex]
 $i$ (degree)                       & $78.77^{+0.11}_{-0.11}$                \\ [1ex]
 $M_1$ ($M_{\sun}$)                 & $0.8252^{+0.0031}_{-0.0027}$           \\ [1ex]
 $M_2$ ($M_{\sun}$)                 & $0.7816^{+0.0027}_{-0.0022}$           \\ [1ex]
 $q \equiv M_2/M_1$                 & $0.9471^{+0.0017}_{-0.0019}$           \\ [1ex]
 $a$ ($R_{\sun}$)                   & $25.733^{+0.030}_{-0.025}$             \\ [1ex]
 $\pi_{\rm orb}$ (mas)              & $20.174^{+0.072}_{-0.072}$             \\ [1ex]
Distance (pc)                      & $49.57^{+0.18}_{-0.18}$    
\enddata
\tablecomments{See Table~\ref{tab:hd27483_results} for the meaning
of the error scale factors $f$, and rms residuals $\sigma$ in the
top portion of the table.}
\end{deluxetable}
\setlength{\tabcolsep}{6pt}

We note that the Gaia DR3 catalog reports a spurious spectroscopic
orbital solution for HD~283882 (class ``SB2'') with a period of
82.8~days, velocity semiamplitudes of 83.4 and 79.9~\kms\ for the
primary and secondary, and an eccentricity of 0.472. The period is not
an integer multiple of the true period of 11.9~days. This appears to
be an instance in which the fully automated Gaia procedures to solve
for a spectroscopic orbit may have been fooled by the sparse coverage
of the observations ($N = 14$), and/or possible confusion in the
assignment of RVs to the primary or secondary, which are not very
dissimilar in brightness. See also \cite{Bashi:2022} and \cite{Rowan:2023}. It
is possible that this problem will be resolved in future Gaia data
releases, once additional observations are included.

HD~283882 is a well-known RS~CVn object, and has been given the
variable star designation V808~Tau. It is an X-ray source, and
displays other typical signs of stellar activity, including emission
cores in the \ion{Ca}{2} H and K lines \citep{Young:1974}, H$\alpha$
emission \citep{Stauffer:1991}, and spot-induced brightness
modulations \citep{Bopp:1980}. The latter authors measured a
peak-to-peak amplitude of 0.1~mag in the Str\"omgren $b$ and $y$
bands, and a rotation period of about 6.8~days, indicating
non-synchronous rotation.
Similar rotational
modulation is seen in the more
recent photometry from TESS, along with frequent flares
(see Section~\ref{sec:activity} below).

\subsection{HD 26874}
\label{sec:hd26874}

The binary nature of HD~26874 was announced by
\cite{Griffin:1981}, who presented a double-lined spectroscopic orbit with a
period of 55.1~days. Out of their 55 observations, about a dozen were
taken at unfavorable phases with severe line blending, and were rejected by them.
We retained 43 measurements for the primary and 41 for the secondary,
made between 1974 and 1980, with relative weights as specified by
the authors. We used their uncertainty of 0.8~\kms\ for an
observation of unit weight in order to compute initial observational errors
for our analysis. \cite{Griffin:1981} pointed out that while HD~26874 lies
above the color-magnitude diagram of the Hyades in combined light,
when properly disentangled, the components fall slightly below it,
suggesting the binary may lie on the far side of the cluster.

As was the case in the previous two systems, HD~26874 was observed
with the CORAVEL instrument, and two RV measures made four days
apart in 1979 were published by \cite{Mermilliod:2009} for both components.
In addition to these, our analysis incorporates 7 observations
(2003--2005) from the public Elodie archive, in which both components
were measured. We assigned them arbitrary initial uncertainties of 0.5~\kms.

HD~26874 was monitored at the CfA with the Digital Speedometers
between 1992 and 2004. The signal-to-noise ratios of the 53 observations
range from 10 to 33 per resolution element, the early ones being
lower. The last visit used a
longer exposure (${\rm SNR} = 57$). The templates adopted for the
cross-correlations with TODCOR have temperatures of 5750 and 5250~K
for the primary and secondary, and $v \sin i$ values of 4 and 6~\kms,
respectively. The spectroscopic flux ratio is $(F_2/F_1)_{\rm sp} = 0.508 \pm
0.018$.

\begin{figure}
\epsscale{1.15}
\plotone{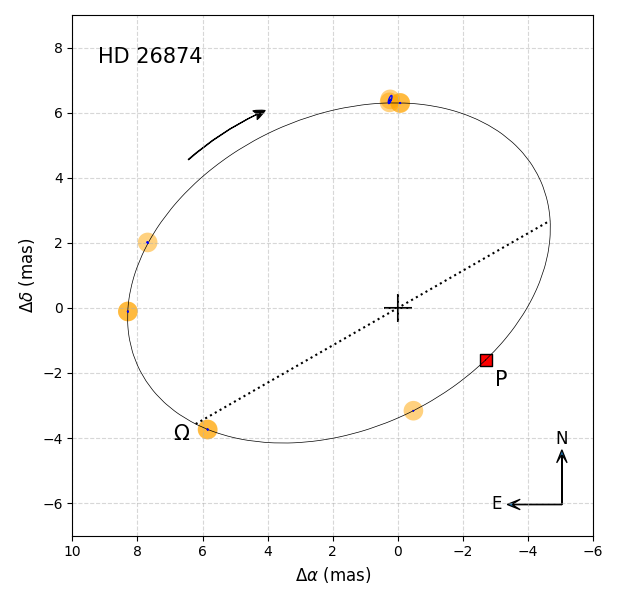}
\figcaption{Astrometric measurements for HD~26874, as in
  Figure~\ref{fig:CHARA_hd27483}.
  \label{fig:CHARA_hd26874}}
\end{figure}

\begin{figure}
\epsscale{1.18}
\plotone{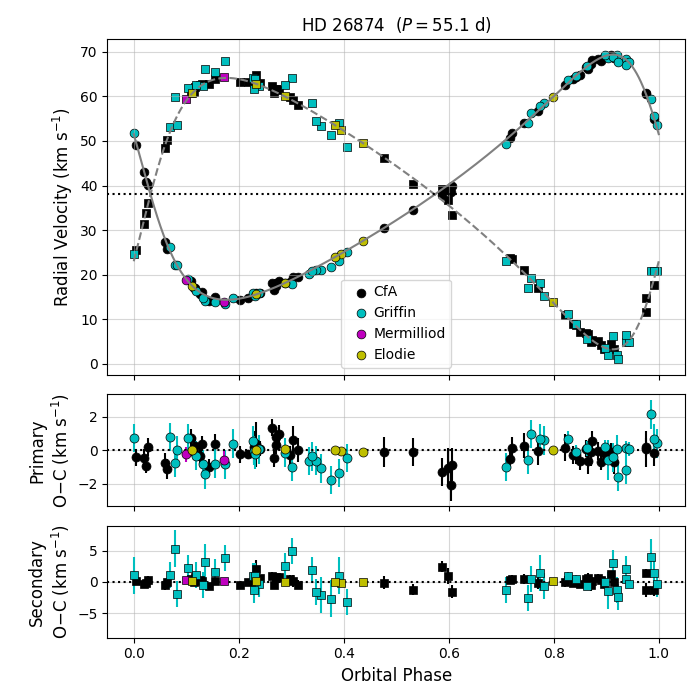}
\figcaption{Same as Figure~\ref{fig:RV_hd27483}, for
  HD~26874.\label{fig:RV_hd26874}}
\end{figure}

The results of our MCMC analysis are given in
Table~\ref{tab:hd26874_results}, together with the systematic offsets we
derived for the various RV data sets relative to ours, as well as the
usual error inflation factors. 
Our orbital parallax for HD~26874 corresponds to a distance of 49~pc,
which is in fact slightly behind the center of the cluster, as
\cite{Griffin:1981} suspected, by about 3.5~pc.
Figure~\ref{fig:CHARA_hd26874}
illustrates the fit to the CHARA observations, and
Figure~\ref{fig:RV_hd26874} shows the RV observations.

\setlength{\tabcolsep}{12pt}
\begin{deluxetable}{lc}
\tablewidth{0pc}
\tablecaption{Results of our Orbital Analysis for HD~26874 \label{tab:hd26874_results}}
\tablehead{
\colhead{~~~~~~~~~~~Parameter~~~~~~~~~~~} &
\colhead{Value}
}
\startdata
 $P$ (day)                          & $55.133176^{+0.000097}_{-0.000090}$  \\ [1ex]
 $T_{\rm peri}$ (HJD$-$2,400,000)   & $50762.584^{+0.014}_{-0.015}$        \\ [1ex]
 $a^{\prime\prime}$ (mas)           & $7.3382^{+0.0032}_{-0.0032}$         \\ [1ex]
 $\sqrt{e}\cos\omega_1$             & $+0.22252^{+0.00054}_{-0.00059}$     \\ [1ex]
 $\sqrt{e}\sin\omega_1$             & $+0.59340^{+0.00037}_{-0.00034}$     \\ [1ex]
 $\cos i$                           & $-0.66385^{+0.00040}_{-0.00037}$     \\ [1ex]
 $\Omega$ (degree)                  & $119.968^{+0.055}_{-0.051}$          \\ [1ex]
 $\gamma$ (\kms)                    & $+38.000^{+0.062}_{-0.062}$          \\ [1ex]
 $K_1$ (\kms)                       & $27.498^{+0.038}_{-0.035}$           \\ [1ex]
 $K_2$ (\kms)                       & $30.436^{+0.048}_{-0.048}$           \\ [1ex]
 $\Delta_{\rm G}$ (\kms)            & $-0.36^{+0.12}_{-0.12}$              \\ [1ex]
 $\Delta_{\rm M}$ (\kms)            & $-0.17^{+0.41}_{-0.27}$              \\ [1ex]
 $\Delta_{\rm E}$ (\kms)            & $+0.309^{+0.066}_{-0.066}$           \\ [1ex]
 $f_{\rm CHARA}$                    & $1.25^{+0.31}_{-0.19}$                \\ [1ex]
 $f_{\rm CfA,1}$ and $\sigma_1$ (\kms) & $1.064^{+0.126}_{-0.099}$, 0.57      \\ [1ex]
 $f_{\rm CfA,2}$ and $\sigma_2$ (\kms) & $0.770^{+0.090}_{-0.071}$, 0.62      \\ [1ex]
 $f_{\rm G,1}$ and $\sigma_1$ (\kms)   & $1.06^{+0.14}_{-0.10}$, 0.71         \\ [1ex]
 $f_{\rm G,2}$ and $\sigma_2$ (\kms)   & $1.01^{+0.13}_{-0.10}$, 1.30      \\ [1ex]
 $f_{\rm M,1}$ and $\sigma_1$ (\kms)   & $1.02^{+1.89}_{-0.50}$, 0.44         \\ [1ex]
 $f_{\rm M,2}$ and $\sigma_2$ (\kms)   & $0.62^{+1.04}_{-0.41}$, 0.25         \\ [1ex]
 $f_{\rm E,1}$ and $\sigma_1$ (\kms)   & $0.105^{+0.054}_{-0.025}$, 0.044     \\ [1ex]
 $f_{\rm E,2}$ and $\sigma_2$ (\kms)   & $0.159^{+0.071}_{-0.037}$, 0.069     \\ [1ex]
\hline \\ [-1.5ex]
\multicolumn{2}{c}{Derived quantities} \\ [1ex]
\hline \\ [-1.5ex]
 $e$                                & $0.40164^{+0.00025}_{-0.00023}$      \\ [1ex]
 $\omega_1$ (degree)                & $69.444^{+0.060}_{-0.056}$           \\ [1ex]
 $i$ (degree)                       & $131.594^{+0.029}_{-0.031}$          \\ [1ex]
 $M_1$ ($M_{\sun}$)                 & $1.0714^{+0.0040}_{-0.0036}$         \\ [1ex]
 $M_2$ ($M_{\sun}$)                 & $0.9682^{+0.0031}_{-0.0031}$         \\ [1ex]
 $q \equiv M_2/M_1$                 & $0.9035^{+0.0021}_{-0.0020}$         \\ [1ex]
 $a$ ($R_{\sun}$)                   & $77.310^{+0.082}_{-0.082}$           \\ [1ex]
 $\pi_{\rm orb}$ (mas)              & $20.411^{+0.026}_{-0.026}$           \\ [1ex]
 Distance (pc)                      & $48.994^{+0.063}_{-0.063}$       
\enddata
\tablecomments{See Table~\ref{tab:hd27483_results} for the meaning
of the error scale factors $f$, and rms residuals $\sigma$ in the
top portion of the table.}
\end{deluxetable}
\setlength{\tabcolsep}{6pt}

The Washington Double Star Catalog \citep[WDS;][]{Worley:1997,
  Mason:2001} has an entry for HD~26874 as WDS~J04157+2049A, and lists
a common proper motion companion some 145\arcsec\ away, which is a
faint and very red star \citep{Reid:1992}.  In a subsequent paper,
\cite{Reid:1997} reported resolving this distant star with HST into two nearly
equal components separated by 0\farcs88 (double star designation RHy\,119). Indeed,
the Gaia DR3 catalog lists both of those red stars with a separation of
0\farcs86 in the N--S direction. They have $G$-band magnitudes of 16.0
and 16.1, more than 8~mag fainter than the primary ($G = 7.65$).
These two faint objects have parallaxes and proper motions that
identify them as members of the cluster, but that are somewhat
different from those of HD~26874.  In particular, the parallaxes are
about 3~mas larger than that of the brighter star. At the distance to
HD~26874, this corresponds to a linear separation of more than 6~pc
along the line of sight, which is large enough to rule out
association.  On the other hand, the two fainter objects may well be
orbiting each other, as their parallaxes are essentially the same and
their proper motions are quite similar.

\subsection{HD 27149}
\label{sec:hd27149}

The first indication that HD~27149 is a spectroscopic binary came from
a report by \cite{Wilson:1948}, who was not able to resolve the
components. He only provided velocities from blended lines, which
nevertheless showed significant variation.  \cite{Woolley:1960} were
the first to successfully measure the primary and secondary
separately. An orbital solution based on 34 measurements for both
stars was later published by \cite{Batten:1973}, giving an orbital
period of 75.6~days and a modest eccentricity of $e = 0.23$. Most of
those observations were taken at relatively low dispersion, and are
poor by today's standards.  The 8 higher dispersion measurements (4.5
and 6.5~\AA~mm$^{-1}$, from 1969--1971) are reasonably good, so we have
chosen to use them here, assigning them an arbitrary initial
uncertainty of 1~\kms.  Five similar quality measurements from
1979--1981 were reported by \cite{McClure:1982}, which we also
incorporated into our modeling, adopting the same initial uncertainty
of 1~\kms.

Additional velocities with much better precision were reported by
\cite{Tomkin:2003}, on the basis of which he significantly improved
the spectroscopic orbital solution. We adopted his 1995--2002
measurements as published, with relative weights as given by the
author and the standard error provided for an observation of unit
weight (0.06~\kms). Three of the 50 epochs were excluded for having
been made when the stars were near conjunction.

Eleven of the 12 CORAVEL measurements from \cite{Mermilliod:2009},
gathered between 1978 and 1992, were incorporated into our analysis as well.
The one excluded was taken when the lines were blended.  A further 7
pairs of RV measurements with the Elodie spectrograph (2003--2005)
were obtained from the public archive, and were assumed initially to
have the same uncertainty of 0.5~\kms\ assigned to them for other systems in this
work.

HD~27149 was observed at the CfA between 1980 and 2001. We collected
71 usable spectra with the Digital Speedometers, at SNRs of 11 to 35
per resolution element. We adopted templates with temperatures of 5750
and 5500~K for the primary and secondary, and rotational broadenings
of 6 and 2~\kms. The primary value is consistent with independent measurements
of 5.5~\kms\ by \cite{Strassmeier:2000} and $7.1 \pm 1.5~\kms$ by
\cite{Mermilliod:2009}. We obtained an estimate of
$(F_2/F_1)_{\rm sp} = 0.610 \pm 0.015$ for the spectroscopic flux ratio at the
mean wavelength of our observations.

This object was observed interferometrically in the $K$ band with
the PTI, between 2000 September and 2008 October.
A total of 100 measurements of the squared visibilities
were incorporated into our analysis. In addition to contributing to
the orbital parameters, they allow an independent estimate of the
flux ratio, $(F_2/F_1)_K$. Because the visibilities are invariant
under a point-symmetric inversion around the binary origin, we restricted
the flux ratio during the analysis to be smaller than unity, as expected
for main-sequence stars with secondaries less massive than the primaries.

\begin{figure}
\epsscale{1.15}
\plotone{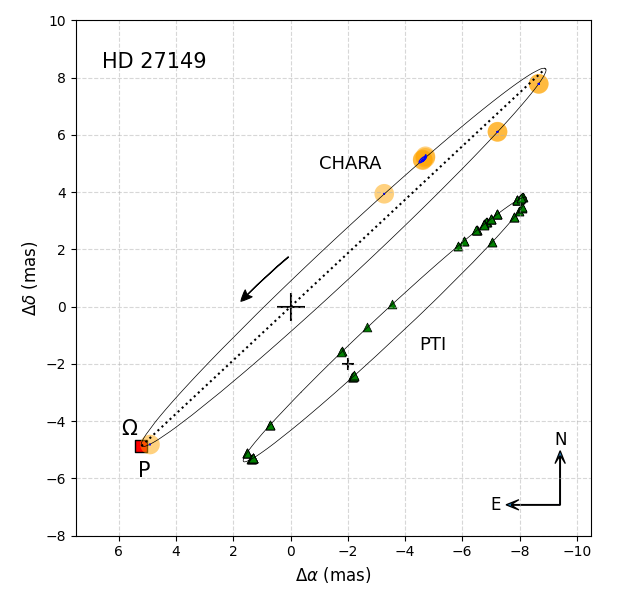}
\figcaption{Astrometric measurements for HD~27149, as in
  Figure~\ref{fig:CHARA_hd27483}. The smaller ellipse is a scaled-down
  version of the orbit meant to show the location of the PTI squared
  visibility measurements (green triangles), which cannot be
  represented directly on the plane of the sky.
  Periastron passage happens to be very close to
  the ascending node in this nearly edge-on orbit.
  \label{fig:CHARA_hd27149}}
\end{figure}

\begin{figure}
\epsscale{1.18}
\plotone{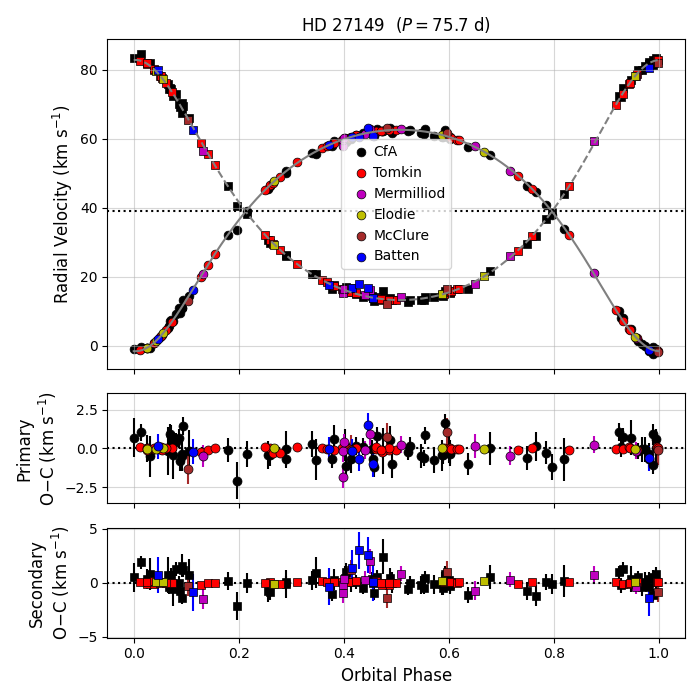}
\figcaption{Same as Figure~\ref{fig:RV_hd27483}, for
  HD~27149.\label{fig:RV_hd27149}}
\end{figure}

The parameters derived from our joint spectroscopic-astrometric
solution are given in Table~\ref{tab:hd27149_results}, with other
derived properties at the bottom. See Figures~\ref{fig:CHARA_hd27149}
and \ref{fig:RV_hd27149} for graphical representations of the astrometric
and RV observations, and the model. The PTI squared visibilities cannot be
represented on the plane of the sky, but we show their phase coverage
in Figure~\ref{fig:CHARA_hd27149}.

\setlength{\tabcolsep}{12pt}
\begin{deluxetable}{lc}
\tablewidth{0pc}
\tablecaption{Results of our Orbital Analysis for HD~27149 \label{tab:hd27149_results}}
\tablehead{
\colhead{~~~~~~~~~~~Parameter~~~~~~~~~~~} &
\colhead{Value}
}
\startdata
 $P$ (day)                          & $75.658178^{+0.000062}_{-0.000062}$  \\ [1ex]
 $T_{\rm peri}$ (HJD$-$2,400,000)   & $50160.429^{+0.012}_{-0.011}$        \\ [1ex]
 $a^{\prime\prime}$ (mas)           & $9.6594^{+0.0023}_{-0.0023}$         \\ [1ex]
 $\sqrt{e}\cos\omega_1$             & $-0.51993^{+0.00011}_{-0.00012}$     \\ [1ex]
 $\sqrt{e}\sin\omega_1$             & $+0.01501^{+0.00067}_{-0.00057}$     \\ [1ex]
 $\cos i$                           & $-0.07204^{+0.00030}_{-0.00030}$     \\ [1ex]
 $\Omega$ (degree)                  & $133.169^{+0.015}_{-0.018}$          \\ [1ex]
 $\gamma$ (\kms)                    & $+38.927^{+0.065}_{-0.060}$          \\ [1ex]
 $K_1$ (\kms)                       & $32.0741^{+0.0084}_{-0.0165}$        \\ [1ex]
 $K_2$ (\kms)                       & $34.770^{+0.015}_{-0.015}$           \\ [1ex]
 $(F_2/F_1)_K$ from PTI              & $0.763^{+0.086}_{-0.027}$            \\ [1ex]
 $\Delta_{\rm T}$ (\kms)            & $+0.462^{+0.062}_{-0.060}$           \\ [1ex]
 $\Delta_{\rm M}$ (\kms)            & $+0.27^{+0.18}_{-0.15}$              \\ [1ex]
 $\Delta_{\rm E}$ (\kms)            & $+0.354^{+0.067}_{-0.062}$           \\ [1ex]
 $\Delta_{\rm Mc}$ (\kms)           & $+0.14^{+0.35}_{-0.32}$              \\ [1ex]
 $\Delta_{\rm B}$ (\kms)            & $+0.83^{+0.25}_{-0.32}$              \\ [1ex]
 $f_{\rm CHARA}$                    & $1.14^{+0.36}_{-0.12}$               \\ [1ex]
 $f_{\rm PTI}$                      & $1.086^{+0.095}_{-0.063}$            \\ [1ex]
 $f_{\rm CfA,1}$ and $\sigma_1$ (\kms) & $0.911^{+0.089}_{-0.070}$, 0.71   \\ [1ex]
 $f_{\rm CfA,2}$ and $\sigma_2$ (\kms) & $0.870^{+0.093}_{-0.057}$, 0.76   \\ [1ex]
 $f_{\rm T,1}$ and $\sigma_1$ (\kms)   & $1.176^{+0.178}_{-0.091}$, 0.079     \\ [1ex]
 $f_{\rm T,2}$ and $\sigma_2$ (\kms)   & $0.900^{+0.134}_{-0.068}$, 0.086  \\ [1ex]
 $f_{\rm M,1}$ and $\sigma_1$ (\kms)   & $1.16^{+0.44}_{-0.14}$, 0.60      \\ [1ex]
 $f_{\rm M,2}$ and $\sigma_2$ (\kms)   & $1.19^{+0.39}_{-0.17}$, 0.77      \\ [1ex]
 $f_{\rm E,1}$ and $\sigma_1$ (\kms)   & $0.057^{+0.044}_{-0.010}$, 0.028  \\ [1ex]
 $f_{\rm E,2}$ and $\sigma_2$ (\kms)   & $0.235^{+0.131}_{-0.041}$, 0.120  \\ [1ex]
 $f_{\rm Mc,1}$ and $\sigma_1$ (\kms)  & $0.80^{+0.59}_{-0.11}$, 0.83      \\ [1ex]
 $f_{\rm Mc,2}$ and $\sigma_2$ (\kms)  & $0.97^{+0.71}_{-0.13}$, 1.01      \\ [1ex]
 $f_{\rm B,1}$ and $\sigma_1$ (\kms)   & $0.713^{+0.375}_{-0.083}$, 0.73      \\ [1ex]
 $f_{\rm B,2}$ and $\sigma_2$ (\kms)   & $1.58^{+0.70}_{-0.24}$, 1.62      \\ [1ex]
\hline \\ [-1.5ex]
\multicolumn{2}{c}{Derived quantities} \\ [1ex]
\hline \\ [-1.5ex]
 $e$                                & $0.26230^{+0.00012}_{-0.00010}$     \\ [1ex]
 $\omega_1$ (degree)                & $178.317^{+0.066}_{-0.072}$         \\ [1ex]
 $i$ (degree)                       & $85.869^{+0.017}_{-0.017}$          \\ [1ex]
 $M_1$ ($M_{\sun}$)                 & $1.1028^{+0.0011}_{-0.0010}$        \\ [1ex]
 $M_2$ ($M_{\sun}$)                 & $1.01736^{+0.00072}_{-0.00109}$     \\ [1ex]
 $q \equiv M_2/M_1$                 & $0.92244^{+0.00044}_{-0.00062}$     \\ [1ex]
 $a$ ($R_{\sun}$)                   & $96.706^{+0.028}_{-0.028}$          \\ [1ex]
 $\pi_{\rm orb}$ (mas)              & $20.4783^{+0.0078}_{-0.0078}$       \\ [1ex]
 Distance (pc)                      & $46.559^{+0.017}_{-0.017}$      
\enddata
\tablecomments{See Table~\ref{tab:hd27483_results} for the meaning
of the error scale factors $f$, and rms residuals $\sigma$ in the
top portion of the table.}
\end{deluxetable}
\setlength{\tabcolsep}{6pt}

\cite{Batten:1973} called attention to the large minimum masses for
the components implied by their spectroscopic solution, which is about
what one would expect from their spectral types. They mentioned the
possibility of eclipses, a suggestion that has been repeated by many other
authors since, and has prompted dedicated photometric efforts to
search for those events \citep[e.g.,][]{Jorgensen:1972}. See a
detailed discussion of those observations by \cite{Tomkin:2003}.
Eclipses were never found, and our interferometric orbit from CHARA and the PTI
now shows that the inclination angle (85\fdg9) is too low for that to
occur. Eclipses would require it to be within about 1\fdg3 of edge-on
\citep{Tomkin:2003}.  The latter author noted that his greatly
improved orbital solution provided the basis for an accurate orbital
parallax for HD~27149, if only the pair could be resolved
interferometrically, such as with the CHARA Array on Mt.\ Wilson that
had just recently begun operations. With the present work, that
promise is finally realized, some 20 years later. The result is indeed very
precise, with a formal parallax uncertainty of only 7.8~$\mu$as (0.04\%), far
better than Gaia DR3, which gives $\pi_{\rm Gaia} = 21.606 \pm 0.042$~mas
(including a zeropoint correction following \citealt{Lindegren:2021b}).
The parallax values themselves do not agree.

As in the case of HD~283882, the Gaia DR3 catalog reports a
spectroscopic orbit for HD~27149 (class ``SB2C'') that is erroneous:
$P = 7.67$~d, $K_1 = 26.6~\kms$, $K_2 = 26.0~\kms$, and $e = 0$. This
is another instance in which the automated orbit-solving algorithms seem
to have failed. And once again, the number of RV measurements is small
(only 13), and the components are relatively similar in brightness.

HD~27149 is also known as an RS~CVn system (V1232~Tau). It was listed by
\cite{Strassmeier:2000} as having \ion{Ca}{2}~H and K emission, and
brightness variations with
a period of about 9 days and an amplitude of 0.05~mag in the
Str\"omgren $y$ band. As shown later, the more recent TESS photometry shows a
somewhat smaller photometric amplitude ($\sim$0.028~mag) at the satellite's
redder bandpass, and a similar though slightly longer
period of 9.7~d. The object is also an X-ray source.

\subsection{HD 30676}
\label{sec:hd30676}

Early RV measurements for HD~30676 include 3 from 1961 by
\cite{Woolley:1965}, one from 1964 by \cite{Kraft:1965}, and a few
by other authors made at much lower dispersion, which have larger
uncertainties. All of these are for the primary component only. While
the first four are quite consistent with our orbit model below, we do
not use them as they provide no significant improvement.

A spectroscopic orbit with a period of about 224.9 days was first
reported by \cite{Mermilliod:2009}, from a series of 39 CORAVEL
observations made between 1978 and 1991. Once again, only the primary
was measured. \cite{Griffin:2012} published an additional 47
observations (1984--2007), in which he was able to measure the faint
secondary component as well. In three of his observations the stellar
lines were blended, and those measures were rejected.

The CfA observations of HD~30676 consist of 27 spectra, gathered with
the Digital Speedometer at the Oak Ridge Observatory in 1992--1997.
The SNRs range between about 10 and 41 per resolution element.
The parameters of the primary
template for the TODCOR analysis were set at $T_{\rm eff} = 6250$~K
and $v \sin i = 12~\kms$. The secondary is too weak to determine those
properties independently from our spectra with any confidence. We
adopted best guesses of $T_{\rm eff} = 5000$~K and $v \sin i =
4~\kms$. The spectroscopic flux ratio is estimated to be $(F_2/F_1)_{\rm sp} =
0.060 \pm 0.012$.

Our MCMC solution used the CfA and CORAVEL RVs, as well as those of Griffin,
with the relative weights he recommended. We initially adopted his
uncertainty for an observation of unit weight of 0.39~\kms.
As pointed out by \cite{Griffin:2012}, the velocity measurements by
\cite{Mermilliod:2009}, which were made without accounting for the
presence of the secondary, are susceptible to ``dragging'' toward
$\gamma$, particularly for the observations near conjunction. Griffin
showed that an orbit for the primary, derived from those measurements
alone, leads to a velocity semiamplitude slightly smaller
than his to a statistically significant degree. Because of the desire
to take advantage of the extended baseline afforded by the CORAVEL
observations, Griffin first used them together with his own in a joint
solution to improve the orbital period, and then carried out a final
fit without them, but with the period held fixed at the previously determined value.

Here we chose a different approach. Instead of fixing the period,
which artificially suppresses correlations with other elements, we
used the CORAVEL, Griffin, and CfA observations together to solve for
all elements, but allowed the CORAVEL measurements for the primary to
have their own velocity semiamplitude $K_{\rm 1,COR}$, so as to not
perturb the $K_1$ value from the others. The remaining elements were
allowed to be shared. The results of this analysis are seen in
Table~\ref{tab:hd30676_results}. Our $K_{\rm 1,COR}$ value is essentially
the same as obtained by Griffin using the CORAVEL velocities alone
($14.81 \pm 0.16~\kms$).

\begin{figure}
\epsscale{1.15}
\plotone{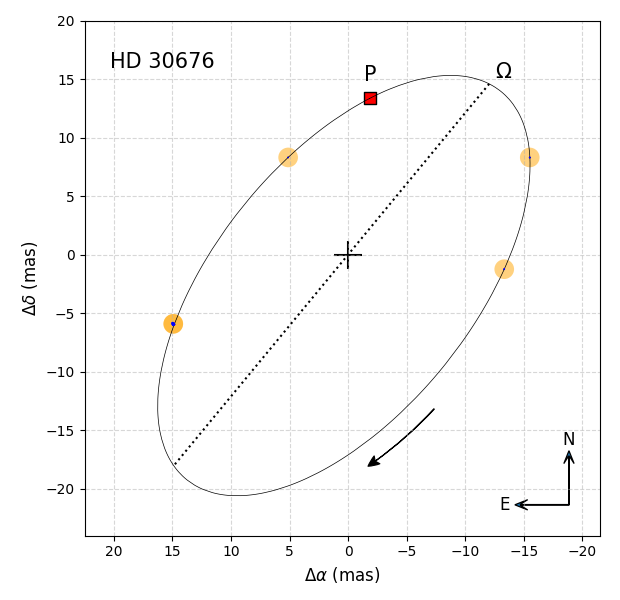}
\figcaption{Astrometric measurements for HD~30676, as in
  Figure~\ref{fig:CHARA_hd27483}.
  \label{fig:CHARA_hd30676}}
\end{figure}

\begin{figure}
\epsscale{1.18}
\plotone{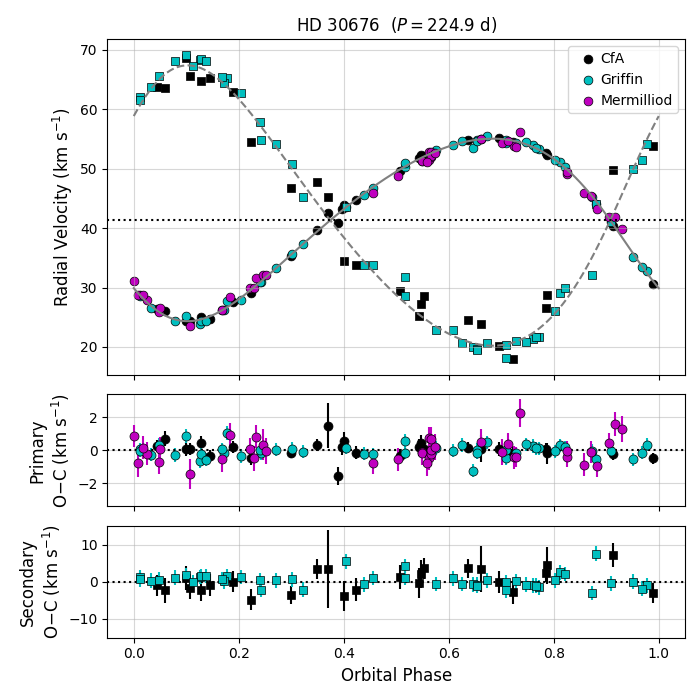}
\figcaption{Same as Figure~\ref{fig:RV_hd27483}, for
  HD~30676.\label{fig:RV_hd30676}}
\end{figure}

In view of our difficulty to determine the best TODCOR template for
the faint secondary, and to guard against potential systematic errors
in the RVs that may come from template mismatch, as a precaution our
analysis included one more free parameter, $\Delta_{\rm CfA}$, corresponding to an overall
shift of our secondary velocities relative to those of the primary.
The solution indicates this offset is not statistically significant.
The astrometric measurements are displayed in
Figure~\ref{fig:CHARA_hd30676}, and the RVs in
Figure~\ref{fig:RV_hd30676}.


\setlength{\tabcolsep}{12pt}
\begin{deluxetable}{lc}
\tablewidth{0pc}
\tablecaption{Results of our Orbital Analysis for HD~30676
\label{tab:hd30676_results}}
\tablehead{
\colhead{~~~~~~~~~~~Parameter~~~~~~~~~~~} &
\colhead{Value}
}
\startdata
 $P$ (day)                          & $224.9214^{+0.0038}_{-0.0039}$  \\ [1ex]
 $T_{\rm peri}$ (HJD$-$2,400,000)   & $50512.61^{+0.13}_{-0.13}$        \\ [1ex]
 $a^{\prime\prime}$ (mas)           & $21.4163^{+0.0084}_{-0.0084}$         \\ [1ex]
 $\sqrt{e}\cos\omega_1$             & $-0.25776^{+0.00040}_{-0.00044}$     \\ [1ex]
 $\sqrt{e}\sin\omega_1$             & $+0.31206^{+0.00056}_{-0.00056}$     \\ [1ex]
 $\cos i$                           & $-0.51238^{+0.00035}_{-0.00035}$     \\ [1ex]
 $\Omega$ (degree)                  & $320.142^{+0.043}_{-0.044}$          \\ [1ex]
 $\gamma$ (\kms)                    & $+41.310^{+0.081}_{-0.081}$          \\ [1ex]
 $K_1$ (\kms)                       & $15.356^{+0.063}_{-0.064}$           \\ [1ex]
 $K_2$ (\kms)                       & $23.58^{+0.34}_{-0.37}$           \\ [1ex]
 $K_{\rm 1,COR}$ (\kms)             & $14.75^{+0.16}_{-0.17}$           \\ [1ex]
 $\Delta_{\rm G}$ (\kms)            & $-0.98^{+0.10}_{-0.10}$              \\ [1ex]
 $\Delta_{\rm M}$ (\kms)            & $-0.17^{+0.15}_{-0.15}$              \\ [1ex]
 $\Delta_{\rm CfA}$ (\kms)          & $-0.59^{+0.63}_{-0.64}$              \\ [1ex]
 $f_{\rm CHARA}$                    & $1.56^{+0.37}_{-0.22}$                \\ [1ex]
 $f_{\rm CfA,1}$ and $\sigma_1$ (\kms) & $1.01^{+0.17}_{-0.12}$, 0.39      \\ [1ex]
 $f_{\rm CfA,2}$ and $\sigma_2$ (\kms) & $0.81^{+0.15}_{-0.10}$, 3.04      \\ [1ex]
 $f_{\rm G,1}$ and $\sigma_1$ (\kms)   & $1.05^{+0.13}_{-0.10}$, 0.40         \\ [1ex]
 $f_{\rm G,2}$ and $\sigma_2$ (\kms)   & $1.46^{+0.18}_{-0.13}$, 1.95      \\ [1ex]
 $f_{\rm M,1}$ and $\sigma_1$ (\kms)   & $1.39^{+0.20}_{-0.14}$, 0.71         \\ [1ex]
\hline \\ [-1.5ex]
\multicolumn{2}{c}{Derived quantities} \\ [1ex]
\hline \\ [-1.5ex]
 $e$                                & $0.16383^{+0.00022}_{-0.00022}$      \\ [1ex]
 $\omega_1$ (degree)                & $129.558^{+0.091}_{-0.091}$           \\ [1ex]
 $i$ (degree)                       & $120.822^{+0.023}_{-0.023}$          \\ [1ex]
 $M_1$ ($M_{\sun}$)                 & $1.262^{+0.042}_{-0.042}$         \\ [1ex]
 $M_2$ ($M_{\sun}$)                 & $0.822^{+0.016}_{-0.016}$         \\ [1ex]
 $q \equiv M_2/M_1$                 & $0.6513^{+0.0107}_{-0.0098}$         \\ [1ex]
 $a$ ($R_{\sun}$)                   & $198.9^{+1.8}_{-1.9}$           \\ [1ex]
 $\pi_{\rm orb}$ (mas)              & $23.16^{+0.22}_{-0.21}$           \\ [1ex]
 Distance (pc)                      & $43.18^{+0.38}_{-0.42}$           \\ [1ex]
 $\Delta G$ (mag)                   & $2.420^{+0.049}_{-0.049}$            \\ [1ex]
 $\Delta Hp$ (mag)                  & $2.39^{+0.92}_{-0.43}$        
\enddata

\tablecomments{See Table~\ref{tab:hd27483_results} for the meaning
of the error scale factors $f$, and rms residuals $\sigma$ in the
top portion of the table.
In this analysis, we have allowed the CORAVEL
  measurements of \cite{Mermilliod:2009} to constrain their own
  velocity semiamplitude for the primary, $K_{\rm 1,COR}$, independent
  of the $K_1$ value from the CfA and \cite{Griffin:2012} observations
  (see the text).}

\end{deluxetable}
\setlength{\tabcolsep}{6pt}

The Gaia DR3 catalog reports an astrometric-spectroscopic orbital
solution for HD~30676, of the kind referred to as ``AstroSpectroSB1''.
Gaia does not resolve the pair, and can only measure the motion of the
center of light. 
This time, the published orbit is correct and the elements are very
similar to ours, though less precise. We reproduce them in
Table~\ref{tab:hd30676_gaia} to facilitate the comparison. The main
difference is a somewhat smaller eccentricity.

\setlength{\tabcolsep}{12pt}
\begin{deluxetable}{lc}
\tablewidth{0pc}
\tablecaption{Orbital Solution for HD~30676 from Gaia DR3 \label{tab:hd30676_gaia}}
\tablehead{
\colhead{~~~~~~~~~~~Parameter~~~~~~~~~~~} &
\colhead{Value}
}
\startdata
 $P$ (day)                                          & $224.78 \pm 0.14$\phn\phn   \\ [1ex]
 $T_{\rm peri}$ (HJD$-$2,400,000)\tablenotemark{a}  & $57360.5 \pm 1.1$\phm{2222} \\ [1ex]
 $a^{\prime\prime}_{\rm phot}$ (mas)                & $6.366 \pm 0.027$           \\ [1ex]
 $e$                                                & $0.1465 \pm 0.0039$         \\ [1ex]
 $\omega_1$ (degree)\tablenotemark{b}               & $131.8 \pm 1.8$\phn\phn     \\ [1ex]
 $i$ (degree)                                       & $120.63 \pm 0.31$\phn\phn   \\ [1ex]
 $\Omega$ (degree)\tablenotemark{b}                 & $320.45 \pm 0.39$\phn\phn   \\ [1ex]
 $\pi_{\rm Gaia}$ (mas)\tablenotemark{c}            & $23.220 \pm 0.023$\phn      \\ [1ex]
 $\gamma$ (\kms)                                    & $+41.384 \pm 0.053$\phn\phs  
\enddata
\tablenotetext{a}{Shifting this value backward by exactly 26 cycles
  using our more precise period yields $T_{\rm peri} = 2,\!451,\!512.8
  \pm 1.1$, in good agreement with our value from
  Table~\ref{tab:hd30676_results}.}
\tablenotetext{b}{We have changed this angle by 180\arcdeg\ because
  Gaia measured the photocenter, which is on the opposite side as the
  secondary relative to the center of mass.}
\tablenotetext{c}{This is the Gaia parallax after accounting for
  orbital motion. Addition of the zeropoint adjustment advocated by
  \cite{Lindegren:2021b} results in the corrected value $\pi_{\rm Gaia}
  = 23.256 \pm 0.023$~mas.}
\end{deluxetable}
\setlength{\tabcolsep}{6pt}

The combination of the semimajor axis of the photocenter from Gaia and
the true relative semimajor axis from CHARA allows us to infer the
brightness difference between the components at the Gaia wavelengths.
This follows from the relation $a_{\rm phot}^{\prime\prime} =
a^{\prime\prime} (B - \beta)$, where $B$ is the secondary's fractional
mass and $\beta$ its fractional brightness. The fractional mass $B =
M_2/(M_1+M_2)$ is expressed in terms of the orbital elements as $B =
K_1/(K_1+K_2)$, and $\beta = (1+10^{0.4 \Delta G})^{-1}$, where
$\Delta G$ is the magnitude difference. We obtained $\Delta G = 2.420
\pm 0.049$~mag.

The Hipparcos catalog \citep{Perryman:1997, vanLeeuwen:2007} 
did not include any information about the orbit, but did report
an acceleration solution for HD~30676, i.e, one that required the
derivatives of the proper motion components, in addition to the 5
standard astrometric parameters, in order to properly model the
motion on the plane of the sky.
A test in which we used the intermediate astrometric data from
the Hipparcos mission (abscissa residuals), together with all other observations,
revealed that the satellite was quite capable of detecting the orbital
motion of the photocenter. We obtained a semimajor axis very similar to
that of Gaia ($a_{\rm phot}^{\prime\prime} = 6.34 \pm 1.03$). However,
the much larger uncertainty makes the correspondingly
uncertain inferred magnitude difference, $\Delta H\!p = 2.39 \pm 0.92$~mag, too
poor to be of much use.

\subsection{HD 28545}
\label{sec:hd28545}

The final object in our sample was first shown to be a spectroscopic binary by
\cite{Griffin:1981}, based on observations taken at two different
telescopes. The secondary was not detected. Their 46 measurements for
the primary star were made between 1972 and 1980, two of which were
rejected for giving large velocity residuals.  Their orbit featured a
modest eccentricity and a period of 358.4~days, which is so close to a
year that a significant gap in their phase coverage was unavoidable.
The only previous observations in the literature seem to be two
measurements by \cite{Wilson:1948}, which did not reveal the duplicity
of the object.

HD~28545 was reobserved at the McDonald Observatory in 1995--2005 by
\cite{Tomkin:2007}, who detected and measured the weak secondary for
the first time. He combined his observations with those of Griffin to
produce an orbital solution with better phase coverage. In the
process, he determined the offsets required for each of Griffin's two
datasets in order to place them on the same velocity system as his.  We adopted
those offsets here as recommended by \cite{Tomkin:2007}. We also used
the relative weights for his observations, and for those of
\cite{Griffin:1981} established by Tomkin in that same combined solution, along
with the corresponding 0.058~\kms\ error for an observation of unit
weight that he reported. Five of the 33 Tomkin observations were affected by blending,
and were excluded.

The velocity of HD~28545 was monitored at the CfA with the Digital
Speedometers from 1979 to 2005. We gathered 75 observations with SNRs
between 9 and 57 per resolution element. Four additional spectra were
obtained with TRES in 2011 and 2012, having SNRs ranging from 50 to
131.  The primary template for the cross-correlations with TODCOR used
$T_{\rm eff} = 5250$~K and $v \sin i = 4~\kms$. As in the case of
HD~30676, we were not able to establish the template parameters for
the faint secondary independently. Our adopted template parameters for
that star were 4500~K and 2~\kms.  To account for possible biases in
our secondary velocities that may come from template mismatch, our
analysis allowed for a systematic offset relative to the RVs of the
primary ($\Delta_{\rm CfA}$), although in the end the offset was insignificant. Our
estimate of the flux ratio from our spectra is $(F_2/F_1)_{\rm sp} = 0.048 \pm
0.009$.

The \cite{Griffin:1981} velocities precede both ours and those of
\cite{Tomkin:2007}, and are therefore useful for constraining the
orbital period. However, because they did not account for the presence
of the unseen secondary, they could be biased toward $\gamma$ and may
affect the semiamplitude $K_1$ to some degree. As done in the previous
section, our analysis allowed those velocities to contribute to all of
the spectroscopic elements except for the semiamplitude, and a
separate $K_{\rm 1,G}$ value was added as a free parameter.
Table~\ref{tab:hd28545_results} shows that the result for this
additional parameter, $K_{\rm 1,G} = 13.81 \pm 0.18~\kms$, is
identical to that of \cite{Griffin:1981}. The difference compared
to the $K_1$ value from the
double-lined observations ($14.028 \pm 0.023~\kms$) goes in the
direction expected for a bias, but is only marginally significant.

\setlength{\tabcolsep}{12pt}
\begin{deluxetable}{lc}
\tablewidth{0pc}
\tablecaption{Results of our Orbital Analysis for HD~28545 \label{tab:hd28545_results}}
\tablehead{
\colhead{~~~~~~~~~~~Parameter~~~~~~~~~~~} &
\colhead{Value}
}
\startdata
 $P$ (day)                           & $358.4369^{+0.0060}_{-0.0060}$  \\ [1ex]
 $T_{\rm peri}$ (HJD$-$2,400,000)    & $49268.91^{+0.15}_{-0.15}$      \\ [1ex]
 $a^{\prime\prime}$ (mas)            & $23.745^{+0.024}_{-0.024}$      \\ [1ex]
 $\sqrt{e}\cos\omega_1$              & $-0.4741^{+0.0014}_{-0.0014}$   \\ [1ex]
 $\sqrt{e}\sin\omega_1$              & $+0.3821^{+0.0023}_{-0.0023}$   \\ [1ex]
 $\cos i$                            & $-0.4607^{+0.0014}_{-0.0014}$   \\ [1ex]
 $\Omega$ (degree)                   & $159.769^{+0.062}_{-0.063}$     \\ [1ex]
 $\gamma$ (\kms)                     & $+40.531^{+0.046}_{-0.042}$     \\ [1ex]
 $K_1$ (\kms)                        & $14.028^{+0.022}_{-0.023}$      \\ [1ex]
 $K_2$ (\kms)                        & $19.872^{+0.049}_{-0.049}$      \\ [1ex]
 $K_{\rm 1,G}$ (\kms)                & $13.81^{+0.18}_{-0.18}$         \\ [1ex]
 $\Delta_{\rm CfA}$ (\kms)           & $+0.11^{+0.32}_{-0.32}$         \\ [1ex]
 $\Delta_{\rm T}$ (\kms)             & $-0.307^{+0.048}_{-0.045}$      \\ [1ex]
 $f_{\rm CHARA}$                     & $0.93^{+0.30}_{-0.16}$           \\ [1ex]
 $f_{\rm TRES,1}$ and $\sigma_1$ (\kms) & $1.21^{+0.86}_{-0.36}$, 0.39    \\ [1ex]
 $f_{\rm TRES,2}$ and $\sigma_2$ (\kms) & $3.17^{+1.96}_{-0.83}$, 2.64    \\ [1ex]
 $f_{\rm DS,1}$ and $\sigma_1$ (\kms)   & $0.764^{+0.073}_{-0.057}$, 0.12 \\ [1ex]
 $f_{\rm DS,2}$ and $\sigma_2$ (\kms)   & $0.564^{+0.055}_{-0.042}$, 3.22 \\ [1ex]
 $f_{\rm T,1}$ and $\sigma_1$ (\kms)    & $1.07^{+0.19}_{-0.13}$, 0.066   \\ [1ex]
 $f_{\rm T,2}$ and $\sigma_2$ (\kms)    & $0.94^{+0.16}_{-0.12}$, 0.190   \\ [1ex]
 $f_{\rm G,1}$ and $\sigma_1$ (\kms)    & $1.13^{+0.15}_{-0.11}$, 0.85    \\ [1ex]
\hline \\ [-1.5ex]
\multicolumn{2}{c}{Derived quantities} \\ [1ex]
\hline \\ [-1.5ex]
 $e$                                 & $0.3708^{+0.0013}_{-0.0013}$    \\ [1ex]
 $\omega_1$ (degree)                 & $141.13^{+0.24}_{-0.24}$        \\ [1ex]
 $i$ (degree)                        & $117.431^{+0.090}_{-0.091}$     \\ [1ex]
 $M_1$ ($M_{\sun}$)                  & $0.9717^{+0.0055}_{-0.0056}$    \\ [1ex]
 $M_2$ ($M_{\sun}$)                  & $0.6859^{+0.0028}_{-0.0028}$    \\ [1ex]
 $q \equiv M_2/M_1$                  & $0.7059^{+0.0022}_{-0.0022}$    \\ [1ex]
 $a$ ($R_{\sun}$)                    & $251.30^{+0.41}_{-0.41}$        \\ [1ex]
 $\pi_{\rm orb}$ (mas)               & $20.318^{+0.042}_{-0.042}$      \\ [1ex]
 Distance (pc)                       & $49.22^{+0.10}_{-0.10}$         \\ [1ex]
 $\Delta G$ (mag)                    & $2.016^{+0.041}_{-0.038}$      
\enddata

\tablecomments{See Table~\ref{tab:hd27483_results} for the meaning
of the error scale factors $f$, and rms residuals $\sigma$ in the
top portion of the table.
In this analysis, we have allowed the measurements of
  \cite{Griffin:1981} to constrain their own velocity semiamplitude
  for the primary, $K_{\rm 1,G}$, independent of the $K_1$ value from
  the CfA and \cite{Tomkin:2007} observations (see the text).
  $\Delta_{\rm CfA}$ is a velocity offset to be added to the secondary
  velocities from CfA to refer them to the same system as the primary
  velocities.}

\end{deluxetable}
\setlength{\tabcolsep}{6pt}

\begin{figure}
\epsscale{1.18}
\plotone{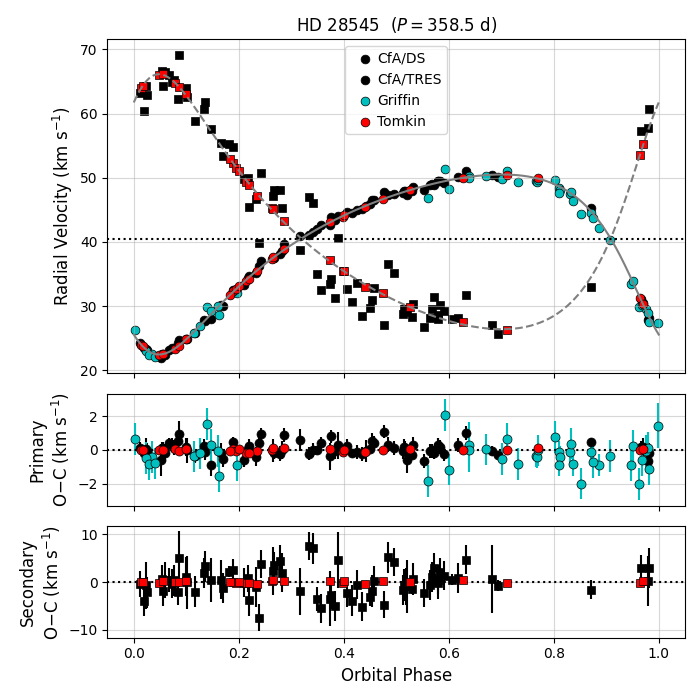}
\figcaption{Same as Figure~\ref{fig:RV_hd27483}, for
  HD~28545.\label{fig:RV_hd28545}}
\end{figure}

Two CORAVEL observations from 1982 by \cite{Mermilliod:2009} are
hardly worth including, as they are only for the primary and may also
be affected by blending. Figure~\ref{fig:RV_hd28545} shows a plot of
the RV observations used for the analysis, along with the model.

\begin{figure}
\epsscale{1.15}
\plotone{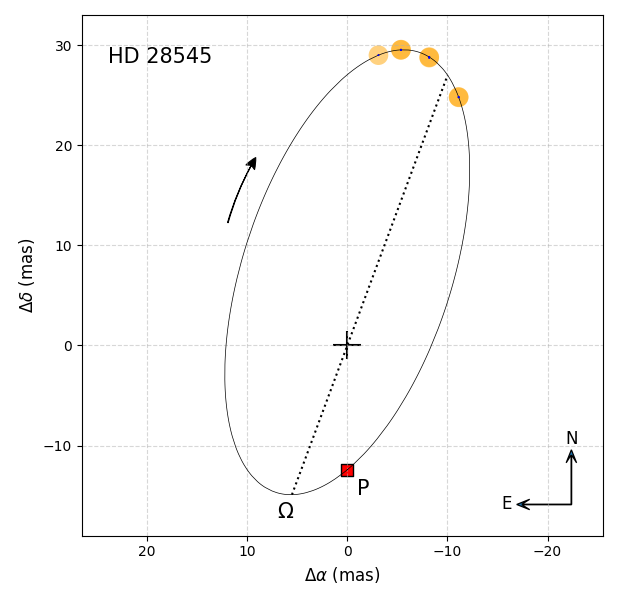}
\figcaption{Astrometric measurements for HD~28545, as in
  Figure~\ref{fig:CHARA_hd27483}. See the text.
  \label{fig:CHARA_hd28545}}
\end{figure}

The CHARA measurements are shown graphically in
Figure~\ref{fig:CHARA_hd28545}. They are seen to all be at one end of
the orbit, which is a consequence of the seasonality of the scheduling
at CHARA and the orbital period being so close to a year ($P =
358.4$~d). However, despite the poor phase coverage, most of the
orbital elements are well determined by the spectroscopy, so the only
astrometric elements that rely solely on the CHARA measurements
($a^{\prime\prime}$, $i$, $\Omega$) are still reasonably well
constrained by virtue of the very high precision achieved with the
MIRC-X and MYSTIC beam combiners (typical separation errors of
$\sim$10\,$\mu$as). A small bias due to the incomplete phase coverage
cannot be excluded, however, and additional measurements will be required
to remove any doubt.

A purely astrometric orbit for HD~28545 has been obtained by the Gaia
mission, and reported as type ``Orbital'' in the DR3 catalog. Once
again the binary was not resolved, and the orbit corresponds to the
photocenter. The elements are essentially correct, but have larger
uncertainties than ours (see Table~\ref{tab:hd28545_gaia}). The main
difference is a smaller eccentricity, as was also the case for
HD~30676.  As in that system, we used the photocentric semimajor axis
and relative semimajor axis to infer the  brightness difference between
the components of HD~28545 in the Gaia bandpass. We obtained $\Delta G
= 2.016 \pm 0.041$~mag.

Our TRES spectra show the \ion{Ca}{2} H and K lines in emission,
indicating the system is chromospherically active. Additional comments
on the activity are presented next.

\setlength{\tabcolsep}{12pt}
\begin{deluxetable}{lc}
\tablewidth{0pc}
\tablecaption{Orbital Solution for HD~28545 from Gaia DR3 \label{tab:hd28545_gaia}}
\tablehead{
\colhead{~~~~~~~~~~~Parameter~~~~~~~~~~~} &
\colhead{Value}
}
\startdata
 $P$ (day)                                          & $357.95 \pm 0.58$\phn\phn   \\ [1ex]
 $T_{\rm peri}$ (HJD$-$2,400,000)\tablenotemark{a}  & $57510.9 \pm 1.7$\phm{2222} \\ [1ex]
 $a^{\prime\prime}_{\rm phot}$ (mas)                & $6.62 \pm 0.10$             \\ [1ex]
 $e$                                                & $0.273 \pm 0.025$           \\ [1ex]
 $\omega_1$ (degree)\tablenotemark{b}               & $139.7 \pm 2.4$\phn\phn     \\ [1ex]
 $i$ (degree)                                       & $118.1 \pm 2.7$\phn\phn     \\ [1ex]
 $\Omega$ (degree)\tablenotemark{b}                 & $160.0 \pm 1.3$\phn\phn     \\ [1ex]
 $\pi_{\rm Gaia}$ (mas)\tablenotemark{c}            & $20.45 \pm 0.26$\phn      
\enddata
\tablenotetext{a}{Shifting this value backward by exactly 23 cycles
  using our more precise period yields $T_{\rm peri} = 2,\!449,\!266.9
  \pm 1.7$, in fair agreement with our value from
  Table~\ref{tab:hd28545_results}.}
\tablenotetext{b}{These angles, as reported by Gaia, are not correct
  for the photocenter. They are flipped by 180\arcdeg\ (which is why
  they agree with ours), probably because there are no Gaia velocities
  with which to establish the correct quadrant.}
\tablenotetext{c}{This is the Gaia parallax after accounting for
  orbital motion. Addition of the zeropoint adjustment advocated by
  \cite{Lindegren:2021b} results in the corrected value $\pi_{\rm Gaia}
  = 20.48 \pm 0.26$~mas.}
\end{deluxetable}
\setlength{\tabcolsep}{6pt}

\section{Stellar activity}
\label{sec:activity}

\begin{figure*}
\centering{
\begin{tabular}{c}
\includegraphics[scale=0.37]{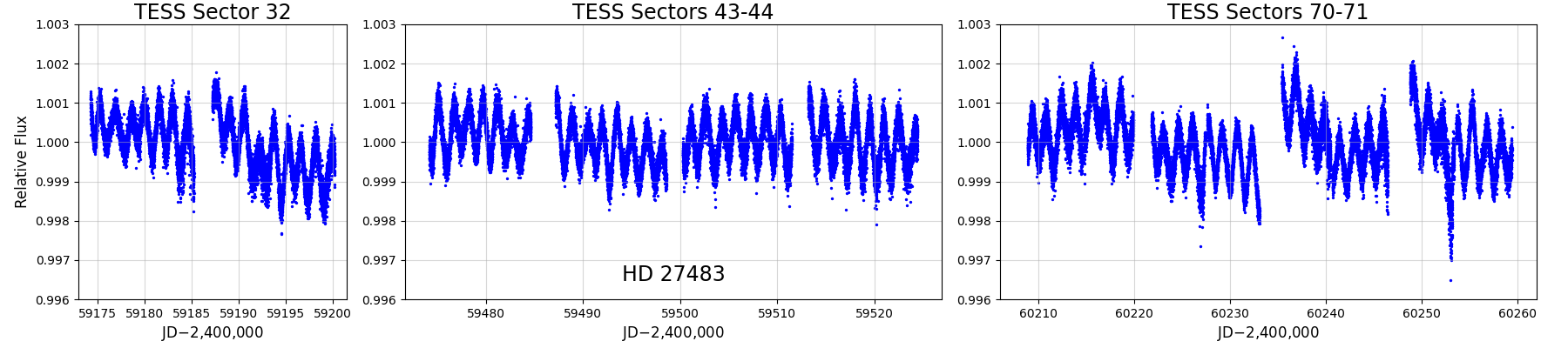} \\
\includegraphics[scale=0.37]{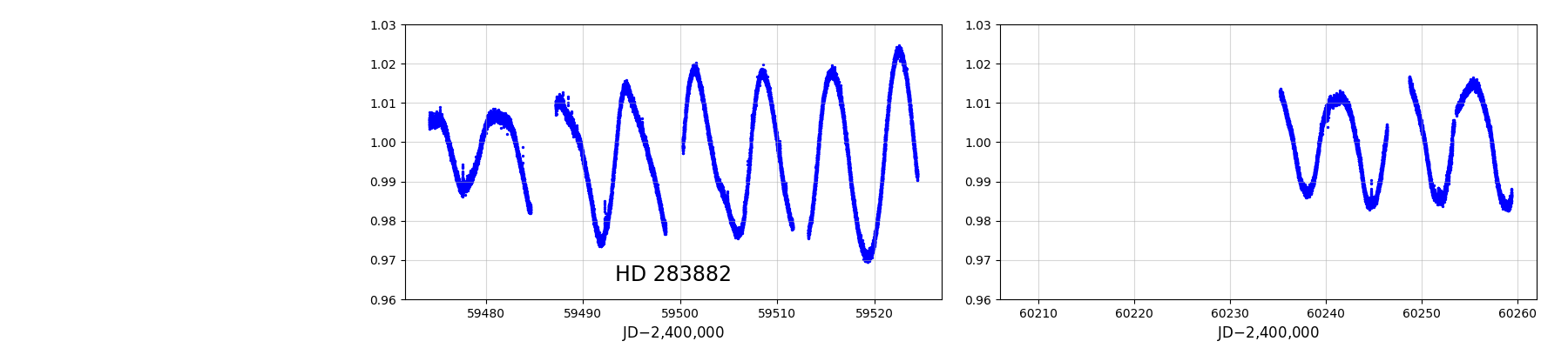} \\
\includegraphics[scale=0.37]{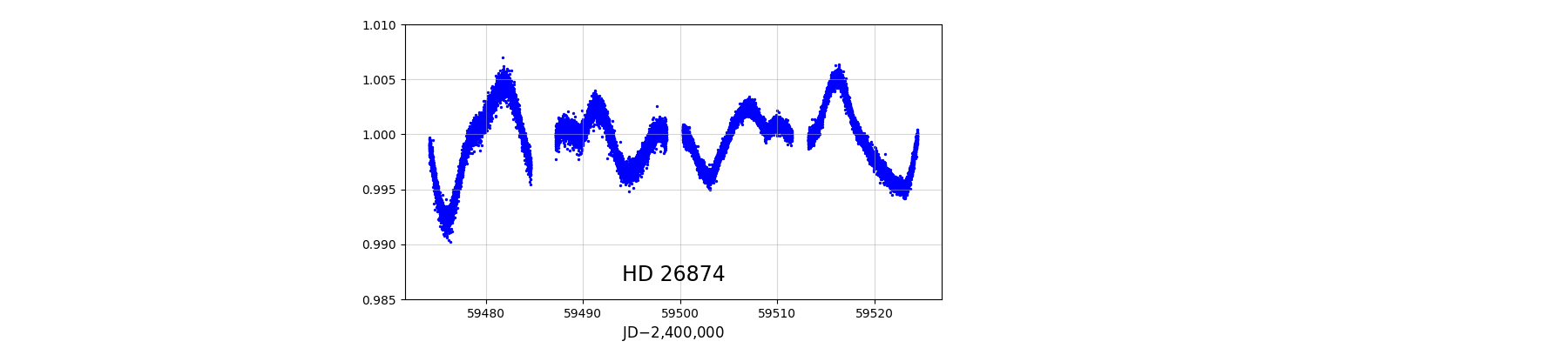} \\
\includegraphics[scale=0.37]{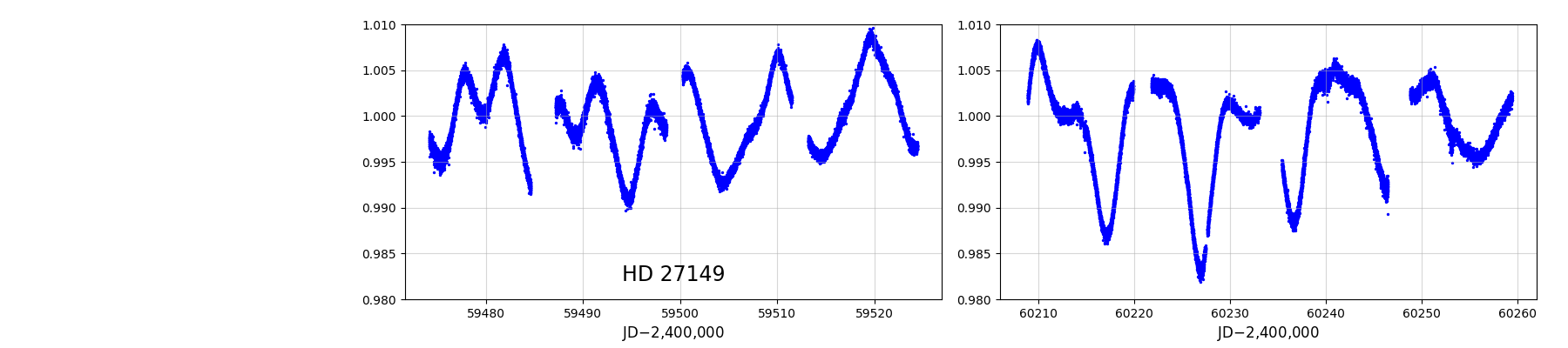} \\
\includegraphics[scale=0.37]{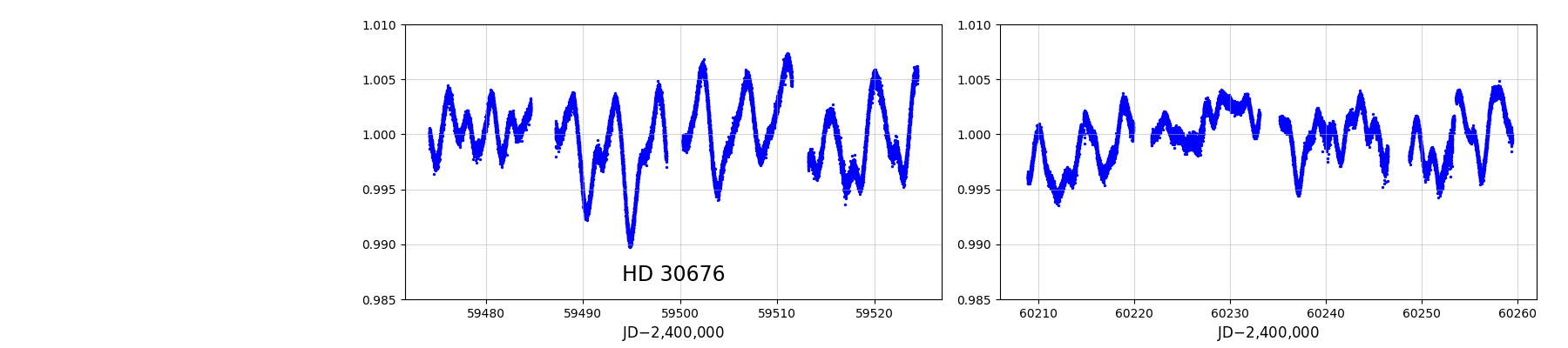} \\
\includegraphics[scale=0.37]{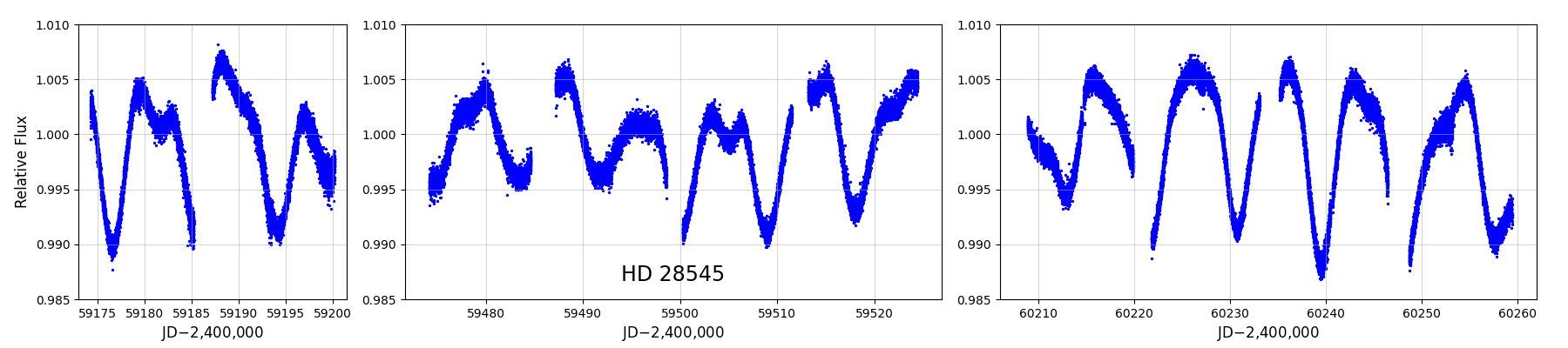}
\end{tabular}
}
\figcaption{TESS light curves (at 2~min cadence) for our six targets,
one per row. Note that the vertical scales are different.\label{fig:tess}}
\end{figure*}

All of our targets have been detected as X-ray sources \citep{Voges:1999,
Voges:2000}, which is a sign of chromospheric activity. The X-ray
luminosities (in erg~s$^{-1}$) range from $\log L_{\rm X} = 29.12$ to 29.64, except for
HD~28545, which is weaker (28.45). Not surprisingly, they
all also display photometric variability, as shown by the
observations from TESS. Figure~\ref{fig:tess} collects the light curves
from the satellite measured at 2~min cadence (simple aperture photometry), which
we downloaded from the Mikulski
Archive for Space Telescopes (MAST)\footnote{\url{https://archive.stsci.edu}}.
For display purposes, we have normalized the fluxes by dividing by
the median in each of the observing sectors, ignoring any long-term trends.
The total amplitudes over the entire observing period range from
less than 5~mmag (HD~27483) to 55~mmag (HD~283882).

Under the common assumption that the variability is caused by spots carried
around as the stars rotate, we have estimated the rotation periods from
the average interval between maxima or minima. They are listed in
Table~\ref{tab:rotation}, along with the peak-to-peak amplitudes. 
The two objects with the largest amplitudes, HD~283882 and HD~27149,
have previously been recognized as RS~CVn systems. In several
cases, the complicated pattern of variation may constitute evidence that
both components of the binary are spotted, or that one or both have more
than one active region, or that the active regions evolve with time. Here we have
focused only on the dominant features in the light curve,
presumably due to the brighter primaries, and derived a single period.

\setlength{\tabcolsep}{8pt}
\begin{deluxetable}{lcc}
\tablewidth{0pc}
\tablecaption{Estimated Rotation Periods of our Targets from
the TESS Photometry. \label{tab:rotation}}
\tablehead{
\colhead{Star} &
\colhead{$P_{\rm rot}$} &
\colhead{Total amplitude}
\\
\colhead{} &
\colhead{(day)} &
\colhead{(mmag)}
}
\startdata
HD 27483   & \phm{\tablenotemark{a}}$1.536 \pm 0.012$\tablenotemark{a}  &  4.5 \\
HD 283882  & $6.942 \pm 0.050$  &  55  \\
HD 26874   & $8.54 \pm 0.48$    &  16  \\
HD 27149   & $9.665 \pm 0.086$  &  28  \\
HD 30676   & $4.361 \pm 0.048$  &  17  \\
HD 28545   & $8.80 \pm 0.10$    &  20  
\enddata
\tablenotetext{a}{This is exactly half of the orbital period.
Rather than rotation, this is most likely the period of the
ellipsoidal variation (see the text).}
\end{deluxetable}
\setlength{\tabcolsep}{6pt}

Within the uncertainty, the period we derive for HD~27483 is exactly
half of the orbital period. The light curve maxima occur consistently at the orbital
quadratures, and the minima at conjunctions. Given the short period
of the binary ($P_{\rm orb} = 3.06$~d), the circular orbit, and the
expectation that the stars' rotations should be synchronized with
the orbital motion at the age of the Hyades, it seems
likely that the photometric changes are caused in whole or at least in part by
the reflection effect (ellipsoidal variation), which has a period of
$P_{\rm orb}/2$. While spots cannot be entirely ruled out, they are
less common in mid-F stars such as these.

HD~283882 has an orbital period long enough ($P_{\rm orb} = 11.93$~d)
that spin-orbit synchronization is not expected. However, it's orbit happens to be the most
eccentric in our sample ($e = 0.518$), and pseudo-synchronization
\citep{Hut:1981} would lead to a rotation period near 4~d. The
value we measure, which is closer to 7~d, would indicate it has not
yet reached that state.

\begin{figure}
\epsscale{1.17}
\plotone{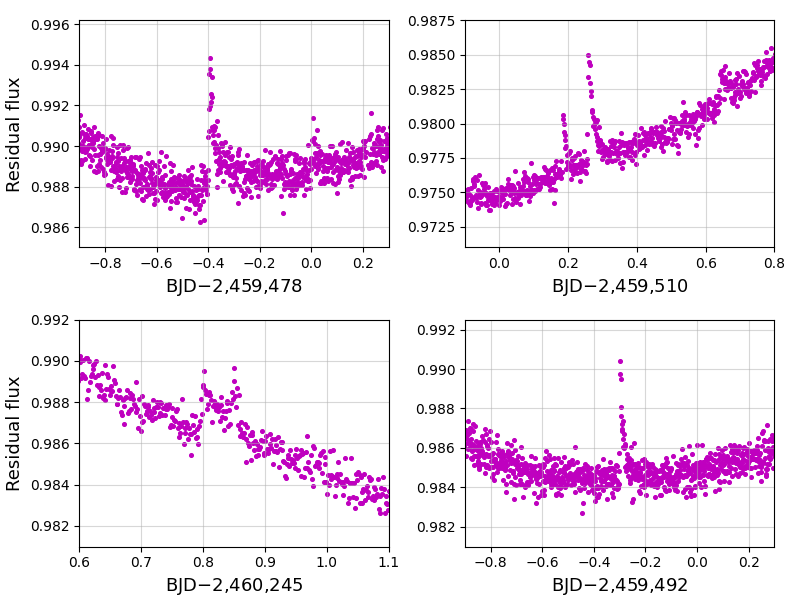}
\figcaption{Examples of flares detected in the TESS light curve of HD~283882.\label{fig:flares}}
\end{figure}

HD~283882 is also the only object in which we detect flares in the TESS light
curves. A few of the more obvious ones are shown in Figure~\ref{fig:flares}.
By visual inspection, we detect approximately 30 of these events over
a total time of observation of about 68 days, not counting gaps. The corresponding
average flare rate is about 0.44 flares per day, although this is
likely to be a lower limit as we have only recorded ones that stand out clearly.
Occasionally these outbursts can occur much more frequently, as seen in
two pairs of events in Figure~\ref{fig:flares} separated by just
one or two hours. It is quite possible that both components are flaring
(spectral types late G or early K).

\section{Discussion}
\label{sec:discussion}

For five of our six targets, we have determined the component masses
with relative uncertainties in the range 0.1--0.7\%. The exception is
HD~30676, which has errors of 3.3\% and 1.9\% for the primary
and secondary, respectively, limited by the spectroscopy. Of all the
previously published dynamical mass determinations in the Hyades, the only
ones with formal errors of about 1\% or better are the eclipsing
system HD~27130 \citep[commonly known also as vB~22;][]{Torres:2002,
  Brogaard:2021}, and HD~284163, which is also part of our CHARA
  program and was reported separately \citep{Torres:2024a}.
We take this opportunity to collect in Table~\ref{tab:previous} all
dynamical mass determinations for the main-sequence components
of non-interacting binaries in the Hyades of which we are aware,
published prior to this work.
They span most of the main sequence in the cluster, from the A7 primary
of $\theta^2$~Tau to the early-to-mid M dwarfs in vA~351~BC.\footnote{The
latter system was overlooked in a few of the previous discussions of the empirical
mass-luminosity relation \citep{Torres:2024a, Torres:2024b}.}

\setlength{\tabcolsep}{8pt}
\begin{deluxetable*}{lcccl}
\tablewidth{0pc}
\tablecaption{Previously Published Dynamical Mass Determinations in the Hyades \label{tab:previous}}
\tablehead{
\colhead{System} &
\colhead{$M_1$} &
\colhead{$M_2$} &
\colhead{$M_3$} &
\colhead{Reference}
\\
\colhead{} &
\colhead{($M_{\sun}$)} &
\colhead{($M_{\sun}$)} &
\colhead{($M_{\sun}$)} &
\colhead{}
}
\startdata
vB 22          &  $1.0245 \pm 0.0024$       &  $0.7426 \pm 0.0016$    &      \nodata       &  \cite{Brogaard:2021}       \\
51 Tau         &  $1.76 \pm 0.08$           &  $1.47 \pm 0.12$        &      \nodata       &  \cite{Anguita-Aguero:2022} \\
$\theta^1$ Tau\tablenotemark{a} &       \nodata              &  $1.28 \pm 0.13$        &      \nodata       &  \cite{Lebreton:2001}       \\
$\theta^2$ Tau &  $2.15 \pm 0.12$           &  $1.87 \pm 0.11$        &      \nodata       &  \cite{Armstrong:2006}      \\
70 Tau         &  $1.363 \pm 0.073$         &  $1.253 \pm 0.075$      &      \nodata       &  \cite{Torres:1997b}         \\
vB 80          &  $1.63^{+0.30}_{-0.13}$    &  $1.11^{+0.21}_{-0.14}$ &      \nodata       &  \cite{Torres:2019a}        \\
HD 28363       &  $1.341^{+0.026}_{-0.024}$ &  $1.210 \pm 0.021$      &  $0.781 \pm 0.014$ &  \cite{Torres:2019b}        \\
vA 351 BC\tablenotemark{b}         &  $0.43 \pm 0.04$        &  $0.41 \pm 0.04$   &  \nodata & \cite{Benedict:2021}       \\
HD 284163      &  $0.784 \pm 0.011$         &  $0.5245 \pm 0.0047$    &  $0.59 \pm 0.12$   &  \cite{Torres:2024a}        \\
vB 120         &  $1.065 \pm 0.018$         &  $1.008 \pm 0.016$      &      \nodata       &  \cite{Torres:2024b}        
\enddata
\tablenotetext{a}{The primary of $\theta^1$~Tau is a giant.}
\tablenotetext{b}{This is a hierarchical quadruple system (AD + BC). The masses listed here
correspond to stars~B and C in the nomenclature of \cite{Benedict:2021}. BC is a close
M-dwarf binary with a period of 0.75~d. Star~A ($0.53 \pm 0.10~M_{\sun}$) is
presumed to be orbited in turn by a white dwarf (star~D) with a mass of
0.54~$M_{\sun}$, and the AD pair orbits BC with a period
of about 2.7~yr. We omit Star~A from this table on the grounds that it may
have interacted with star~D in the past. We also do not list the K dwarf primary in
V471~Tau, an eclipsing post-common envelope system in the Hyades \citep[see, e.g.,][]{Muirhead:2022},
because of the prior interaction with its white dwarf companion.}
\end{deluxetable*}
\setlength{\tabcolsep}{6pt}

Most earlier mass determinations in the Hyades have been compared
against stellar evolution models in a diagram of absolute $V$
magnitude vs.\ mass \citep[see, e.g.,][and references therein]{Torres:2019b}.
The spectroscopic flux ratios reported in this work allow us to add
six more systems to that diagram, and our CHARA observations
now enable a comparison in the near infrared as well ($H$, $K$).

The flux ratios from our spectra correspond strictly to wavelengths
centered on the region of the \ion{Mg}{1}~b triplet ($\sim$5187~\AA).
We have converted them to the more standard $V$ band, by employing
PHOENIX model spectra from \cite{Husser:2013} to interpolate the flux
ratios between 5187~\AA\ and the $H$ band, using appropriate
temperatures for the binary components. We list these $V$-band ratios in
Table~\ref{tab:photometry}, along with those in $H$ and $K$ from
CHARA, averaged over all observations for each binary. We note that
although it is less precise, the $K$-band flux ratio from the PTI for
HD~27149 (Table~\ref{tab:hd27149_results}),
agrees perfectly with the result from CHARA.
Also included
in Table~\ref{tab:photometry} are the apparent $V$ magnitudes of each object, taken from the
homogeneous catalog of \cite{Mermilliod:1991}, and the $H$ and $K$
magnitudes from 2MASS \citep{Cutri:2003}. We used these, together with
the orbital parallaxes from this work, to compute absolute magnitudes.
Extinction is negligible for the Hyades.

\setlength{\tabcolsep}{4pt}
\begin{deluxetable*}{lccccccc}
\tablewidth{0pc}
\tablecaption{Combined-Light Photometry, Flux Ratios, and Orbital
Parallaxes for our Targets \label{tab:photometry}}
\tablehead{
\colhead{Target} &
\colhead{$\pi_{\rm orb}$} &
\colhead{$V$} &
\colhead{$H$} &
\colhead{$K_S$} &
\colhead{$(F_2/F_1)_V$} &
\colhead{$(F_2/F_1)_H$} &
\colhead{$(F_2/F_1)_K$}
\\
\colhead{} &
\colhead{(mas)} &
\colhead{(mag)} &
\colhead{(mag)} &
\colhead{(mag)} &
\colhead{} &
\colhead{} &
\colhead{}
}
\startdata
 HD 27483   &  $21.174 \pm 0.073$\phn    &  $6.173 \pm 0.017$  &  $5.155 \pm 0.024$  &  $5.062 \pm 0.018$  &  $0.87 \pm 0.03$      &  $0.932 \pm 0.002$  &  $0.930 \pm 0.003$  \\
 HD 283882  &  $20.174 \pm 0.072$\phn    &  $9.555 \pm 0.031$  &  $7.120 \pm 0.038$  &  $6.956 \pm 0.038$  &  $0.61 \pm 0.04$      &  $0.783 \pm 0.039$  &  $0.811 \pm 0.097$  \\
 HD 26874   &  $20.411 \pm 0.026$\phn    &  $7.835 \pm 0.005$  &  $6.257 \pm 0.017$  &  $6.190 \pm 0.017$  &  $0.55 \pm 0.02$      &  $0.719 \pm 0.024$  &  $0.698 \pm 0.002$  \\
 HD 27149   &  $21.4783 \pm 0.0078$\phn  &  $7.528 \pm 0.011$  &  $6.032 \pm 0.016$  &  $5.950 \pm 0.017$  &  $0.64 \pm 0.02$      &  $0.742 \pm 0.016$  &  $0.763 \pm 0.009$  \\
 HD 30676   &  $23.16 \pm  0.22$\phn     &  $7.114 \pm 0.005$  &  $5.725 \pm 0.023$  &  $5.666 \pm 0.020$  &  $0.0850 \pm 0.0012$  &  $0.233 \pm 0.009$  &  $0.249 \pm 0.009$  \\
 HD 28545   &  $20.318 \pm 0.042$\phn    &  $8.938 \pm 0.009$  &  $6.902 \pm 0.021$  &  $6.817 \pm 0.018$  &  $0.0795 \pm 0.0010$  &  $0.306 \pm 0.009$  &  $0.337 \pm 0.025$  
\enddata
\end{deluxetable*}
\setlength{\tabcolsep}{6pt}

The formal uncertainties for the flux ratios of HD~283882
in all three bandpasses are significantly larger than for the other
five targets. The same is true of the $V$, $H$, and $K$ magnitudes for
the combined light. This extra scatter is caused by the high level of
activity of the system, mentioned earlier.
The masses and absolute magnitudes for the individual binary
components of our six targets are collected in Table~\ref{tab:magnitudes}.

\setlength{\tabcolsep}{3pt}
\begin{deluxetable*}{lcccccccc}
\tablewidth{0pc}
\tablecaption{Masses and Individual Absolute Magnitudes for our
Targets \label{tab:magnitudes}}
\tablehead{
\colhead{Target} &
\colhead{$M_1$} &
\colhead{$M_2$} &
\colhead{$M_{V_1}$} &
\colhead{$M_{V_2}$} &
\colhead{$M_{H_1}$} &
\colhead{$M_{H_2}$} &
\colhead{$M_{K_1}$} &
\colhead{$M_{K_2}$}
\\
\colhead{} &
\colhead{($M_{\sun}$)} &
\colhead{($M_{\sun}$)} &
\colhead{(mag)} &
\colhead{(mag)} &
\colhead{(mag)} &
\colhead{(mag)} &
\colhead{(mag)} &
\colhead{(mag)}
}
\startdata
 HD 27483  &  $1.363 \pm 0.010$   &  $1.3323 \pm 0.0099$   &  $3.482 \pm 0.025$ &  $3.633 \pm 0.027$ &  $2.499 \pm 0.025$ &  $2.575 \pm 0.025$ &  $2.405 \pm 0.020$ &  $2.484 \pm 0.020$ \\
 HD 283882 &  $0.8252 \pm 0.0029$ &  $0.7816 \pm 0.0025$   &  $6.596 \pm 0.042$ &  $7.133 \pm 0.055$ &  $4.272 \pm 0.045$ &  $4.537 \pm 0.049$ &  $4.125 \pm 0.070$ &  $4.352 \pm 0.084$ \\
 HD 26874  &  $1.0714 \pm 0.0038$ &  $0.9682 \pm 0.0031$   &  $4.860 \pm 0.015$ &  $5.509 \pm 0.026$ &  $3.395 \pm 0.023$ &  $3.753 \pm 0.027$ &  $3.314 \pm 0.017$ &  $3.705 \pm 0.017$ \\
 HD 27149  &  $1.1028 \pm 0.0011$ &  $1.01736 \pm 0.00091$ &  $4.725 \pm 0.017$ &  $5.210 \pm 0.023$ &  $3.295 \pm 0.019$ &  $3.619 \pm 0.021$ &  $3.226 \pm 0.018$ &  $3.519 \pm 0.019$ \\
 HD 30676  &  $1.262  \pm 0.042$  &  $0.822 \pm  0.016$    &  $4.026 \pm 0.006$ &  $6.703 \pm 0.015$ &  $2.776 \pm 0.024$ &  $4.358 \pm 0.041$ &  $2.731 \pm 0.021$ &  $4.241 \pm 0.037$ \\
 HD 28545  &  $0.9717 \pm 0.0056$ &  $0.6859 \pm 0.0028$   &  $5.560 \pm 0.010$ &  $8.310 \pm 0.016$ &  $3.731 \pm 0.023$ &  $5.017 \pm 0.033$ &  $3.672 \pm 0.028$ &  $4.853 \pm 0.064$ 
\enddata
\end{deluxetable*}
\setlength{\tabcolsep}{6pt}

\begin{figure}
\epsscale{1.17}
\plotone{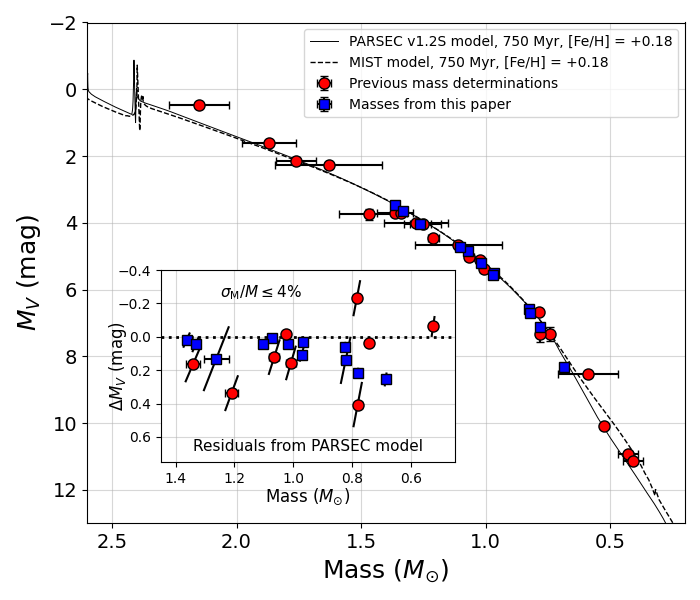}
\figcaption{Empirical $V$-band mass-luminosity relation in the Hyades
cluster, compared against two model isochrones, as labeled. The inset
shows the difference in absolute magnitude from the PARSEC 1.2S model,
restricted to stars with relative mass errors no larger than 4\%.
The differences in the sense observation minus model are predominantly positive,
indicating theory overestimates the visual flux.
Diagonal lines in the inset represent the
deviation in the absolute magnitudes, factoring in the
correlation with the mass errors (see the text). \label{fig:MLR_V}}
\end{figure}

Figure~\ref{fig:MLR_V} shows the empirical mass-luminosity relation
for the cluster in the visual band, and includes the 10 systems in
Table~\ref{tab:previous} and the 6 new ones from this work. The
measurements are compared against two model isochrones with the
same age and metallicity for the Hyades \citep[750~Myr,
${\rm [Fe/H]} = +0.18$;][]{Brandt:2015, Dutra-Ferreira:2016}.
These properties were adopted here for consistency with other recent
studies reporting dynamical mass determinations in the Hyades
\citep{Torres:2019a, Torres:2019b, Torres:2024a, Torres:2024b}.
Overall, the isochrones follow the observed trend quite well, and there is very
little difference between the PARSEC~1.2S model of \cite{Chen:2014}
and the MIST model of \cite{Choi:2016}. The most significant difference
between the curves is at the low-mass end. Here, the models from the PARSEC
series include an ad hoc adjustment to the temperature-opacity relation below
$T_{\rm eff} = 4730$~K ($\sim$0.7~$M_{\sun}$), which was introduced
by the builders to
better reproduce the measured radii for M dwarfs in eclipsing binaries.
\cite{Chen:2014} argued that this also improves
the fits in the color-magnitude diagrams of clusters.

A closer look at the stars with the more precise mass measurements
reveals that the models seem to overestimate their brightness
by a small amount. A similar
observation has been made previously by \cite{Torres:2024b}, for the
vB~120 system. The inset in Figure~\ref{fig:MLR_V} shows this more clearly.
It plots the $M_V$ deviations (observed minus predicted) from the
PARSEC~1.2S isochrone, for all stars with relative mass errors $\sigma_M/M$
smaller than 4\%, which includes all six binary systems in this paper.
Note that because of the slope of the mass-luminosity relation, the
deviation $\Delta M_V$ from the model for any given star is different at
$M + \sigma_M$ than at $M - \sigma_M$, resulting in the diagonal error bars
in the figure. The same type of comparison for the MIST model
gives marginally larger deviations, in the same direction.

In principle, a different choice for the age and/or metallicity of the
models could reduce the discrepancy with theory.
The effect of those properties on the model predictions was discussed
in detail by \cite{Torres:2024b}, who noted that age estimates
for the Hyades have typically ranged between 625 and 800~Myr, and
metallicity estimates from about ${\rm [Fe/H]} = +0.1$ to +0.2.
The values adopted here are near the high end in both cases.
Lowering the metallicity makes the models brighter, which would
produce even larger differences between the models and the observations.
Reducing the age, on the other hand, has a negligible effect 
on the brightness of stars of these masses because they are still
unevolved at these ages. A systematic error in the observations
themselves seems implausible, given that the various binary studies are largely
independent. We conclude, therefore, that the deviations are real,
and that based on this sample of 21 stars between about 0.5 and
1.4~$M_{\sun}$, the PARSEC and MIST isochrones overestimate
the visual flux by $0.096 \pm 0.029$~mag over this mass range, on average.
A possible explanation is missing opacities in the models.

\begin{figure}
\epsscale{1.17}
\plotone{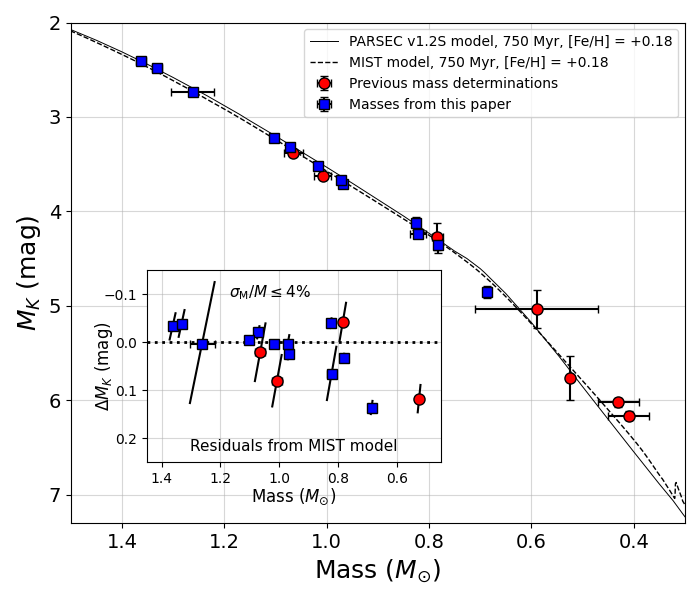}
\figcaption{Similar to Figure~\ref{fig:MLR_V}, for the $K$ band.
In this case, the MIST model is a better fit to the observations.
\label{fig:MLR_K}}
\end{figure}

The situation changes toward the near infrared. Figure~\ref{fig:MLR_K}
presents the empirical mass-luminosity in the (2MASS) $K$ band, where fewer systems
with dynamical masses and known component fluxes are available.
At this wavelength, the MIST model is fainter than the PARSEC model
by about 0.04~mag, and provides a better fit to the observations. The
inset displays the brightness deviations from the MIST isochrone in the
same way as before. The average $\Delta M_K$ difference from 16 stars
($0.020 \pm 0.013$~mag) is now much smaller, and barely significant.
The conclusion we draw is then that the MIST model is able to give a satisfactory
match to the observations in $K$, at least over the mass range
0.5--1.4~$M_{\sun}$. The PARSEC model still overpredicts
the $K$-band fluxes by a small amount, as noted above.

In the $H$ band we find a result that is intermediate between $V$
and $K$ (not shown). As above, the MIST isochrone is fainter than the
one from PARSEC, and provides a better match to the observed fluxes,
but still predicts the stars should be brighter. The average difference in $M_H$ is
$0.062 \pm 0.017$~mag, based on 14 stars over the same mass range as before.

\begin{figure}
\epsscale{1.17}
\plotone{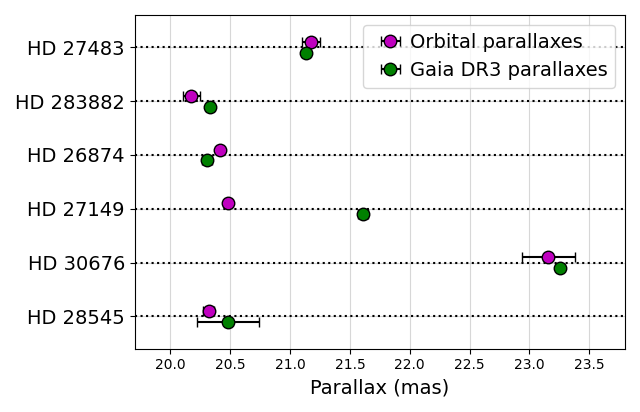}
\figcaption{Comparison between the Gaia DR3 parallaxes (with the
zeropoint corrections of \citealt{Lindegren:2021b}) and the orbital
parallaxes determined in this work.\label{fig:px}}
\end{figure}

A comparison between the orbital parallaxes of our six targets and those
from Gaia is seen in Figure~\ref{fig:px}. For HD~30676 and HD~28545,
the Gaia parallaxes we adopt are the ones corrected for orbital motion, as listed
in Tables~\ref{tab:hd30676_gaia} and \ref{tab:hd28545_gaia},
respectively. In all cases except for one, our parallaxes and those from Gaia agree to
within 2$\sigma$ or better. The outlier is HD~27149, which happens
to have the most precise orbital parallax in our sample
($\sigma_{\pi} = 7.8~\mu$as). The reasons for the large discrepancy are
unclear, but may be related to the large RUWE value from Gaia for this
object (${\rm RUWE} = 1.936$). Motion in the 75.6 day orbit was not
accounted for in Gaia's astrometric solution, as it was for the
longer-period binaries in our sample, and the much shorter
spectroscopic period inferred by the mission is incorrect, as mentioned
in Section~\ref{sec:hd27149}. We estimate the amplitude (semimajor axis)
of the motion of the photocenter of HD~27149 to be about 1.3~mas, and we speculate
this could have affected the Gaia parallax.

Under the assumption that all Hyades members share a common space motion,
the expected velocity of a star along the line of sight can be
calculated from the cluster's space velocity vector and the sky position of the
object. The predicted RVs for our 6 binary targets, based on the known space
motion of the Hyades \citep{Gaia:2018}, are in good
agreement with the measured center-of-mass velocities for 5 of them,
with deviations $\gamma - RV_{\rm pred}$ ranging from $-0.24$ to
$+0.57~\kms$. These differences are similar to, or just slightly
larger than the velocity dispersion of the cluster adopted in the Gaia
study mentioned above ($\sigma_{\rm cl} = 0.40~\kms$). The
deviation for HD~27483, on the other hand, is about 2.5 times larger
than $\sigma_{\rm cl}$ in absolute value, or $-1.05~\kms$. While an offset as
large as this is not ruled out by statistics, we note that
HD~27483 also happens to be the only target in our sample that is
known to be in a triple system (see Section~\ref{sec:hd27483}),
and this could be affecting its measured $\gamma$ velocity. Indeed,
over the time interval covered by our observations, the expected RV
of the binary in its $\sim$184~yr orbit around the white dwarf companion,
based on the preliminary elements of \cite{Zhang:2023}, is about
$-0.6~\kms$ from the center of mass of the triple. This accounts for
a large fraction of the negative offset we find, bringing HD~27483
within a more comfortable 0.4 or 0.5~\kms\ of the expected velocity.

\section{Conclusions}
\label{sec:conclusions}

Using the CHARA Array, and in one case the PTI, we have spatially resolved
6 double-lined  spectroscopic binaries in the Hyades. We have combined those
observations with new and existing radial-velocity measurements,
to derive high-precision dynamical masses for all components,
in addition to the orbital parallaxes. When adding the determinations
for HD~284163 reported separately \citep{Torres:2024a}, our CHARA
program has significantly increased the number of non-interacting
systems with mass determinations in the cluster, from 9 to 16 (see
Table~\ref{tab:previous}).

Five of the 6 targets in this work have yielded masses with relative
errors better than 1\%, which are among the best in the Hyades to date.
In one case (HD~27149), the uncertainty in our orbital
parallax is only 7.8~$\mu$as (0.04\%).

We have compared all systems with mass determinations in the cluster
against two different models of stellar evolution (PARSEC~1.2S, and MIST).
Isochrones calculated for an age and metallicity fixed to values
appropriate for the Hyades follow the general trend in the empirical
mass-luminosity relation relatively well in the $V$, $H$, and $K$
bandpasses, over much of the main sequence. This test is valuable because
it involves no free parameters in the comparison with the models.
However, a closer look at the systems with the best mass measurements
(0.5--1.4~$M_{\sun}$) reveals that model fluxes are slightly
overestimated in the visual band over this range,
by about 0.1~mag for the PARSEC isochrone. This may be due at least
in part to missing opacities in the models. The deviation is marginally
larger for MIST. We point out that it is unlikely that these disagreements can be
eliminated with a different choice for the age and/or metallicity in the
models, within reason.
We find that the discrepancy is reduced toward the near infrared: in
the $H$ band it is about 0.06~mag for the MIST models, which perform better
than PARSEC, and in the $K$ band the disagreement largely disappears for
MIST, whereas a small difference remains for the PARSEC model.

All 6 of our targets display evidence of stellar activity, in the
form of X-ray emission, and photometric variability caused by spots rotating
in and out of view. We have estimated their rotation periods using
the light curves from TESS. For HD~283882 we detected
numerous flares, and estimate an average flaring rate of 0.44 events per day.

\begin{acknowledgements}

The spectroscopic observations at the CfA were obtained by
P.\ Berlind,
M.\ Calkins, 
J.\ Caruso, 
R.\ Davis,
G.\ Esquerdo,
J.\ Peters,
E.\ Horine, and
J.\ Zajac.
We are grateful for their assistance.

We also thank R.\ J.\ Davis and J.\ Mink for maintaining the databases of
echelle spectra, and the anonymous referee for a helpful suggestion.
This work is based upon observations obtained with the Georgia State
University Center for High Angular Resolution Astronomy Array at Mount
Wilson Observatory. The CHARA array is supported by the National
Science Foundation under grant No.\ AST-1636624 and AST-2034336.
Institutional support has been provided from the GSU College of Arts
and Sciences and the GSU Office of the Vice President for Research and
Economic Development. MIRC-X received funding from the European
Research Council (ERC) under the European s Horizon 2020 research and
innovation program (grant No. 639889).  J.D.M. acknowledges funding
for the development of MIRC-X (NASA-XRP NNX16AD43G, NSF-AST 1909165)
and MYSTIC (NSF-ATI 1506540, NSF-AST 1909165). Time at the CHARA array
was granted through the NOIRLab community access program (NOIRLab
PropID: 2020B-0010, 2021B-0008, 2022B-235883; PI: G.\ Torres). This
research has made use of the Jean-Marie Mariotti Center Aspro and
SearchCal services. S.K.\ and C.L.D.\ acknowledge support by the
European Research Council (ERC Starting grant, No.\ 639889 and ERC
Consolidator grant, No.\ 101003096), and STFC Consolidated Grant
(ST/V000721/1).  A.L.\ received funding from STFC studentship
No.\ 630008203. 
The Palomar Testbed Interferometer was operated by the NASA Exoplanet Science Institute and the PTI collaboration. It was developed by the Jet Propoulsion Laboratory, California Institute of Technology with funding provided from the National Aeronautics and Space Administration.

This research has benefited from the use of the SIMBAD and VizieR
databases, operated at the CDS, Strasbourg, France, of NASA's
Astrophysics Data System Abstract Service, and of the WEBDA database,
operated at the Department of Theoretical Physics and Astrophysics of
the Masaryk University (Czech Republic).  The work has also made use
of data from the European Space Agency (ESA) mission Gaia
(\url{https://www.cosmos.esa.int/gaia}), processed by the Gaia Data
Processing and Analysis Consortium (DPAC,
\url{https://www.cosmos.esa.int/web/gaia/dpac/consortium}). Funding
for the DPAC has been provided by national institutions, in particular
the institutions participating in the Gaia Multilateral Agreement. The
computational resources used for this research include the Smithsonian
High Performance Cluster (SI/HPC), Smithsonian Institution
(\url{https://doi.org/10.25572/SIHPC}).

\end{acknowledgements}


\end{document}